\documentclass[submitting]{nst}

\usepackage{subfigure,dcolumn}

\usepackage[T2A,T1]{fontenc}
\usepackage[russian,english]{babel}
\usepackage{upgreek}

\usepackage{listings}
\usepackage{graphicx}
\usepackage{comment}
\usepackage{multirow}

\begin{document}

\title{A High-Granularity Proton CT Enhanced by Track Discrimination}
\thanks{This research was supported by the School of Physics, Zhejiang University.}

\author{Huang-Chao Shi}
\affiliation{School of Physics, Zhejiang University, Hangzhou 310058, China}
\author{Dan-Qi Wang}
\affiliation{School of Physics, Zhejiang University, Hangzhou 310058, China}
\author{Mei-Qi Liu}
\affiliation{School of Physical Sciences, University of Science and Technology of China, Hefei 230026, China}
\author{Yi-Dong Yang}
\affiliation{School of Physical Sciences, University of Science and Technology of China, Hefei 230026, China}
\author{Bao-Hua Qi}
\affiliation{Institute of High Energy Physics, Chinese Academy of Sciences, Beijing 100049, China}
\affiliation{University of Chinese Academy of Sciences, Beijing 100049, China}
\author{Yong Liu}
\affiliation{Institute of High Energy Physics, Chinese Academy of Sciences, Beijing 100049, China}
\affiliation{University of Chinese Academy of Sciences, Beijing 100049, China}
\author{Yang-Hui Qin}
\affiliation{School of Nuclear Science and Technology, University of Science and Technology of China, Hefei 230026, China}
\affiliation{State Key Laboratory of Particle Detection and Electronics, University of Science and Technology of China, Hefei 230026, China}
\author{Chang-Qing Feng}
\affiliation{State Key Laboratory of Particle Detection and Electronics, University of Science and Technology of China, Hefei 230026, China}
\affiliation{Department of Modern Physics, University of Science and Technology of China, Hefei 230026, China}
\author{Yi Liu}
\affiliation{School of Physics, Zhengzhou University, Zhengzhou 450001, China}
\author{Hong-Bo Zhu}
\email[Corresponding author, ]{hongbo_zhu@zju.edu.cn}
\affiliation{School of Physics, Zhejiang University, Hangzhou 310058, China}

\begin{abstract}
Proton Computed Tomography (pCT) provides a promising solution to enhance the accuracy of Relative Stopping Power (RSP) required for proton therapy planning. 
This research introduces a novel high-granularity pCT architecture that incorporates a silicon pixel tracking system and a calorimetric range telescope, which uniquely integrates range telescope functionality with track discrimination capabilities.
The Bortfeld function fitting and Convolutional Neural Network (CNN) classifier algorithms are developed and applied for discrimination.
In simulation studies, both approaches demonstrate the capability to reduce uncertainty in Water Equivalent Path Length (WEPL) determination for individual proton tracks to below 3~mm. 
The standard imaging protocol (3.2~mGy, $4\times10^{8}$ protons) achieves sub-millimeter spatial resolution ($\sim$0.5~mm) with sub-1\% RSP accuracy.
With proton count requirements reduced by track discrimination, an ultra-low-dose protocol (0.16~mGy, $2\times10^{7}$ protons) is proposed with achieved sub-1\% RSP accuracy and $\sim$1.1~mm spatial resolution in simulation. 
This low-dose performance significantly expands clinical applicability, particularly for pediatric imaging or frequent imaging scenarios. 
Furthermore, the target 10 MHz proton detection rate suggests potential for real-time image guidance during radiotherapy.
By circumventing the need for ultra-precise energy measurements, this design minimizes hardware complexity and provides a scalable foundation for future pCT systems.
\end{abstract}

\keywords{Proton computed tomography, Range telescope, Track discrimination, Monte Carlo simulation}
\maketitle

\nolinenumbers

\section{Introduction}\label{sec:Intro}
Proton and heavy ion therapies have gained significant recognition for their efficacy in tumor treatment due to the unique Bragg peak feature~\cite{Review_2022_PTCurrentAndFuture}. 
This physical characteristic enables concentrated radiation deposition within tumor volumes while sparing surrounding healthy tissues. 
Precise beam positioning, angular alignment, and energy modulation during treatment planning are crucial for ensuring accurate dose delivery. 
The Relative Stopping Power (RSP) map serves as the fundamental bridge connecting these phases, traditionally derived from Hounsfield Units (HU) in X-ray CT through empirical HU-to-RSP calibration~\cite{HURSP_1998_Calibration}. 
However, inherent uncertainties of 2.4 – 5\% in this conversion~\cite{HURSP_2012_StoichimetricCalib} have motivated active exploration of advanced modalities, including proton CT (pCT) and dual-energy CT (DECT), to improve RSP accuracy.

Current pCT implementations employ two distinct methodologies for RSP mapping: proton-integrating and proton-tracking. 
The integrating approach measures aggregate energy deposition from proton ensembles, while the tracking method reconstructs individual proton tracks and residual energies. 
Although proton-integrating systems benefit from simplified detector architectures and rapid imaging capabilities~\cite{EarlyStudy_1999_RadGraphy}, their spatial resolution is fundamentally limited by Multiple Coulomb Scattering (MCS)~\cite{MCS_1994}, along with high dose (cGy-level) requirements due to proton pileup~\cite{pRGpCT_2013_DoseMeasure}.
In contrast, proton-tracking systems achieve superior spatial resolution ($\sim$1~mm) and significant lower dose (mGy-level) by reconstructing individual proton tracks. 
Due to advancements in detector technologies and high-speed electronics~\cite{Advances_Det_DAQ_2023}, the clinical viability of proton tracking systems has significantly enhanced in recent years~\cite{pCTreview_2015, pCTreview_2018}.  
It should be noted that clinical pCT requires 3D RSP mapping with a density resolution better than 1\% and a spatial resolution better than 1~mm. Such goals are empirical  achieved through detection of more than $10^{4}$ protons per mm$^{2}$~\cite{pCTreview_2018}. 
To ensure clinical practicality, these systems must maintain sustained proton detection rates exceeding 10~MHz to reduce scanning time to the tens of seconds range.

%-----pCT projects
Proton tracking systems typically consist of tracking detectors that measure the positions and directions of both incoming and outgoing protons, followed by range or energy detectors that measure the residual energies of the outgoing protons. 
Silicon-based detectors are extensively used in tracking applications due to their excellent spatial resolution, implemented either as strip detectors~\cite{PhaseII_2016_Operation, PRaVDA_2018_update, iMPACT_2018_design, Bergen_2020_Optimization2} or pixel detectors~\cite{iMPACT_2018_design, Bergen_2020_Optimization2}. 
The strategies for range or energy detection, however, vary significantly across different system designs. 
The Phase-II scanner~\cite{PhaseII_2016_NovelScintDesign} utilized a 5-stage scintillator-photomultiplier tube (PMT) stack to measure proton energy~\cite{PhaseII_2016_TechAndFirstExp}. 
The PRaVDA~\cite{PRaVDA_2018_update}, Hi'CT~\cite{IMPCAS_2023_HiCT-ConceDesign} and Bergen~\cite{Bergen_2020_Optimization2} systems all employed alternating silicon-absorber layers to determine proton range. 
Meanwhile, iMPACT~\cite{iMPACT_2019_CaloPrototype} and ASTRA~\cite{pCTDesign_2022_NRTconcept} used high-granularity calorimeters to simultaneously measure proton ranges and resolve multi-proton events, aided by fine longitudinal and lateral segmentation in the detectors. 
Although both architectures rely on energy deposition data, their implementations differ fundamentally. 
In the iMPACT calorimeter design, dual-threshold energy measurements were employed to identify the buildup region and sub-peak position of the proton energy deposition profiles, thus serving as a dedicated range telescope. 
In contrast, the ASTRA design combined calorimetric energy measurement with range estimation to reconstruct the residual proton energy, thereby functioning as a dual-purpose range telescope and calorimeter.

This study presents an innovative pCT design that integrates a silicon pixel tracking system and a high-granularity calorimetric range telescope. 
Following the iMPACT paradigm, our architecture fundamentally redefines the calorimeter as a range telescope rather than a precision energy spectrometer. 
This approach significantly alleviates the stringent demands for ultra-precise energy measurement. 
However, the range telescope exhibits compromised energy-range correlation when protons undergo nuclear interaction or large-angle scattering.
To address this limitation, the acquired energy deposition profile is novelly transformed into a track classifier through systematic analysis of the proton interaction dynamics and assisted with Bortfeld function fitting and Convolutional Neural Network (CNN) techniques.
Dedicated algorithms have been developed for proton track reconstruction and discrimination. 
The performance of the pCT system has been thoroughly verified through Monte Carlo simulations, demonstrating effective integration of range telescope with energy measurement.
In addition, a simplified prototype is under development to demonstrate the viability of the detector and electronics components.

\section{Materials and Methods}\label{sec:Mat}
%=======================
%    detector design      
%=======================
\subsection{pCT System Design}
\label{pCTDesign}
\begin{figure}[!htb]
	\includegraphics[width=0.95\hsize]{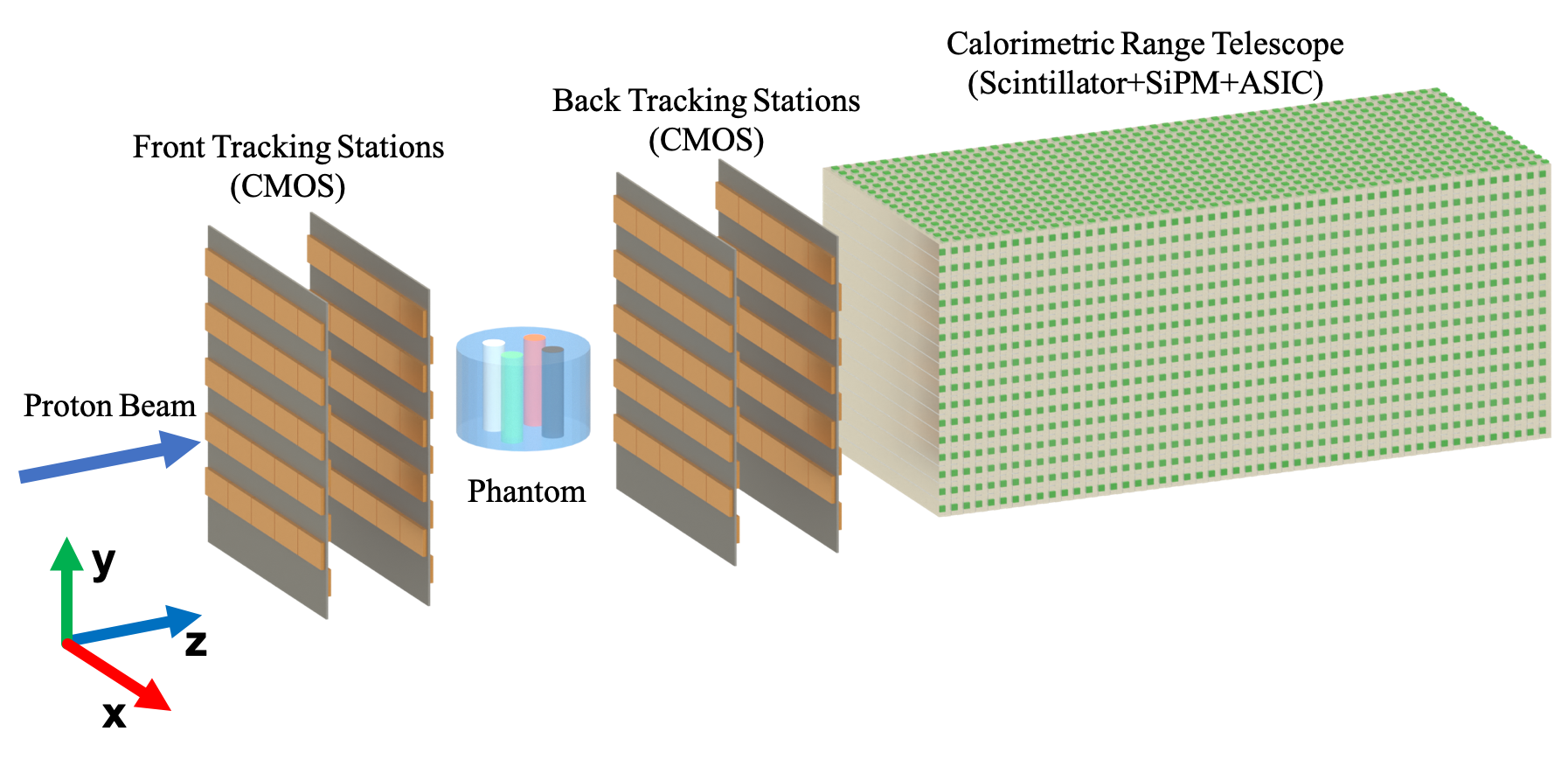}
	\includegraphics[width=0.95\hsize]{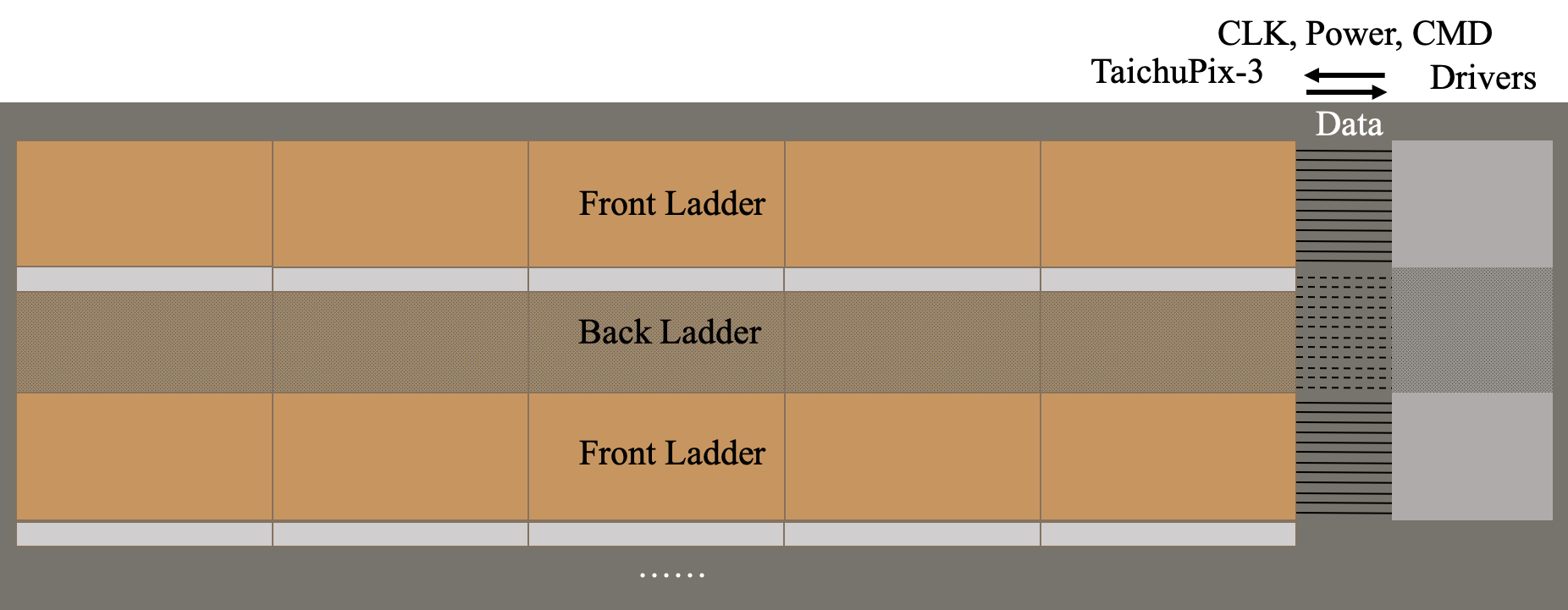}
	\caption{
	The top figure illustrates an overview of pCT system. Each tracking station comprises 5$\times$10 TaichuPix-3 chips (brown), mounted on FPCs and attached to a carbon fiber plate (dark grey). The range telescope consists of 100 layers, each containing 16 scintillator strips (silver) coupled to SiPMs (green).
	The bottom figure shows the ladder structure. The insensitive region (light grey) is occupied by readout electronics; however, the detection efficiency is maintained by overlapping sensitive areas (brown) of the adjacent ladder on the reverse side. 
	}
	\label{fig:pCT_structure}
\end{figure}

The pCT system, as illustrated in Fig.~\ref{fig:pCT_structure} (top), consists of a tracking system and a high-granularity calorimetric range telescope. 
It is designed to handle 200~MeV protons at a maximum rate of 10~MHz across a 100~$\times$~100~mm${^2}$ detection area. 
While the current configuration is tailored for head scans, the system allow for future expansion to accommodate both larger fields of view and greater object thickness.
To reduce lateral proton leakage, the detector coverage is extended to 128~$\times$~128~mm${^2}$. 
The target imaging performance aims for a RSP accuracy of <1\% and a spatial resolution of $\sim$1~mm.
Within this detector coordinate system, the $z-$axis corresponds to the direction of the incoming proton beam. In the transverse plane, the $x-$axis is oriented horizontally perpendicular to the beam ($z-$axis), while the $y-$axis points vertically. For tomographic imaging, the phantom rotates around the $y-$axis during CT scans. 

%====tracker=====
The tracking system includes four tracking stations that are instrumented with Complementary Metal Oxide Semiconductor (CMOS) pixel sensors. 
The two front stations measure positions and directions of protons entering the phantom, while the two back stations measure exiting protons. 
The utilized CMOS pixel sensor, named TaichuPix-3~\cite{TaichuPix1_2021_design, TaichuPix2_2022_Development}, is a full-reticle-sized sensor featuring small pixels of $25~\times~25~\upmu$m$^2$. 
Its high granularity and short dead time (< 500~ns) enable it to achieve high counting efficiency under extreme beam conditions, with the capacity to handle up to 3.6~$\times$~10$^{7}$~hits/(cm$^{2}\cdot$s)~\cite{TaichuPix3_2024_BeamTest}. 
These exceptional performance characteristics ensure that the tracking system can process the target proton fluxes without saturation. %10~MHz without saturation. 
The TaichuPix-3 sensor features a pixel array of 1024 columns and 512 rows, forming an active area of 12.8~$\times$~25.6~mm$^2$. 
Through modular assembly, five sensors are mounted sequentially on a flexible printed circuit (FPC) to form a ladder measuring 12.8~$\times$~128.0~mm$^2$, as shown in Fig.~\ref{fig:pCT_structure} (bottom). 
Ten such ladders are attached to both sides of a 400~$\upmu$m-thick carbon fiber plate, with five ladders on each side. 
They are interleaved with appropriate overlap to minimize the dead area occupied by the peripheral readout electronics. 
This arrangement provides each tracking station with a full detection area coverage of 128~$\times$~128 mm$^2$, while maintaining a material budget of approximately 0.37\% radiation length ($X_{0}$).

%====scintillator=====
The calorimetric range telescope comprises 100 stacked layers, each with a thickness of 3~mm, aligned along the beam axis to determine proton stopping positions.  
Each layer is divided into 16 independent scintillator bars. 
The orthogonal orientation of scintillator bars in adjacent layers enables distinct reconstruction of proton tracks in the $x-y$ plane through cross-layer position correlation. 
This configuration also allows the estimation of proton entry points into the scintillator, facilitating the matching of proton tracks reconstructed from the downstream stations of the tracking system. 
Each scintillator bar measures 128~$\times$~8~$\times$~3~mm$^3$, resulting in a total coverage area of 128~$\times$~128~mm$^2$ and a total thickness of 300~mm. 
BC-408 plastic scintillators are adopted to detect proton energy deposition due to its advantages of short decay time (< 2.1~ns), high light yield and cost-effectiveness. 
The calculated penetration depth of proton beam in BC-408 is 261.2~mm, which is well within the longitudinal containment capacity of the range telescope. 
The scintillator bars are encapsulated with ESR reflector film to increase the light yield and aluminum foil to minimize the inter-channel crosstalk. 
Each scintillator bar is coupled to a Hamamatsu SiPM S13360-3025PE and read out with ASIC MPT2321, which features 32-channels with integrated 12-bit ADC and 20-bit TDC for each individual channel~\cite{MPT_2024_SiPMReadout}. 
The range telescope attains an expected detection rate beyond 10~MHz per channel, enabled by SiPM decay time below 100~ns and MPT2321 integration time of 50~ns.
Data from MPT2321 readout chips and TaichuPix-3 pixel sensors are routed through a FELIX-based DAQ system~\cite{Felix_2019}, achieving sustained throughput up to 10~Gb/s for real-time event packaging and PC transmission.

%=======================
%   Simulation 
%=======================
\subsection{pCT Simulation}
A Geant4 simulation with 200 MeV parallel proton beam was developed to evaluate the system performance and compare designed configurations.
The physics list \texttt{QGSP\_BIC} was employed for proton and secondary particle interactions, while \texttt{G4EmStandardPhysics\_option4} and \texttt{G4OpticalPhysics} governed scintillation photon generation and transport.

Each tracking station comprised TaichuPix-3 chips, FPCs, support, and interconnects, resulting in a material budget of 0.37\% $X_{0}$. 
Noise and crosstalk were intentionally excluded in this preliminary modeling, allowing direct recording of all proton and secondary particle hits without energy thresholds.
In the range telescope, scintillator bars were encapsulated with an 80~$\upmu$m-thick composite reflective layer (ESR + aluminum) as insensitive region. 
To account for light quenching, Birk's attenuation in Geant4 is incorporated to modify the effective energy deposition.

Scintillation light transport was decoupled from the main simulation through a standalone single-bar modeling, with a light yield of 8000~photons/MeV.
The photo-electron yield ($n_{\rm PE}$) was determined from the simulated photon count at the SiPM, then applying a 40\% quantum efficiency and introducing a 10\% smearing to account for coupling uncertainties.  
The $n_{\rm PE}$ exhibited strong correlation with proton hit position along scintillator length (128~mm), varying by 50\% from one side to the other. 
The position dependence was modeled by $n_{\rm PE}$ distributions according to the hit position.
During full-system simulation, light yields were determined by sampling from these position-dependent distributions.

Two phantoms were built in Geant4 to systematically evaluate the pCT system performance and reconstruction algorithm: 
one dedicated to RSP imaging, and the other for Water Equivalent Path Length (WEPL) calibration and validation, where WEPL represents the integral of RSP along the proton path.
\begin{itemize}
    \item \textbf{Phantom 1 (RSP Characterization)}: 
    A cylindrical water phantom (30~mm radius and 40~mm height) containing four cylindrical inserts (5~mm radius and 40~mm height) of polypropylene (PP), teflon, air, and bone-equivalent material (bone-100). The CT scan employed 1$^\circ$ rotational increments over 0--180$^\circ$ for RSP imaging.
    \item \textbf{Phantom 2 (WEPL Calibration)}: 
    A water cuboid (128~$\times$~128~mm$^2$ base area) with adjustable thickness (0--140~mm in 5~mm steps).
    This configuration directly calibrates WEPL from proton residual range measurement by matching the thickness to WEPL.
\end{itemize}

The simulation captured both detector outputs and ground truth data. 
The detector outputs comprised triggered cells in tracking system and $n_{\rm PE}$ map in range telescope.
The ground truth data includes proton energy loss in the phantom and interaction features, whose impact on proton track concluded as follows~\cite{ProPhys_2015_ProTheraphy}:
\begin{itemize}
\item \textbf{Ionization}: Induces quasi-continuous energy loss.
\item \textbf{Elastic Coulomb scattering}: Causes track deviations through scattering.
\item \textbf{Non-elastic nuclear reactions}: Generates secondary particles (protons, neutrons, heavy ions and gamma rays) accompanied by abrupt energy loss in secondary protons.
\item \textbf{Bremsstrahlung}: Simultaneously alters track and induces radiative energy loss, negligible for 200 MeV protons.
\end{itemize}

The range telescope configuration prioritizes a strong proton energy-range correlation.
Thus, only ionization and small-angle elastic Coulomb scattering are considered beneficial for reconstruction.
To facilitate the development of noise suppression algorithms, the simulation recorded binary flags for occurrences of nuclear reactions and maximum scattering angles in both phantom and range telescope, respectively. 

%=======================
%   Reconstruction
%=======================
\subsection{pCT Reconstruction}
The RSP map is derived from detector outputs via a three-step workflow:
\begin{itemize}
\item \textbf{Proton track reconstruction}:
Proton tracks through the tracking system and range telescope are reconstructed to determine both positions and WEPL values, which are derived from residual range measurement with calibrated WEPL-range conversion.
\item \textbf{WEPL imaging}: 
WEPL maps are generated through statistical accumulation of sufficient proton events.
\item \textbf{RSP imaging}: 
The 3D RSP map is reconstructed from multi-angle WEPL projections using the filtered back-projection (FBP) algorithm.
\end{itemize}
The current reconstruction framework predominantly addresses single-proton events with an emphasis on noise suppression, whereas the extension to multi-proton reconstruction remains under development.

\subsubsection{Position Measurement and Filter on Scattering Angle}
\label{sec:Rec_TrackingStation}
The fired pixels are matched between the front and back tracking stations to reconstruct the incoming and outgoing proton tracks, respectively. 
A matching algorithm similar to that presented in Ref.~\cite{pCTDesign_2022_NRTconcept} is used for future multi-proton reconstruction.
In the current single-proton scenario, this algorithm demonstrates sufficient robustness to effectively exclude false matching from secondary particles.

Once matched tracks are established, the Most Likely Path (MLP) algorithm is applied to estimate the track's spatial coordinates in both the x- and y-directions. 
The MLP framework leverages a matrix-based formalism \cite{MLP_2008_MLPFormalism}, which combines Bayesian statistical inference with a Gaussian approximation model of MCS. 
Inputs to the scattering geometry algorithm include predefined phantom boundaries, though these may be replaced in clinical implementations by fixed patient position or a priori imaging data.
To benchmark the MLP algorithm's localization improvement, the inter-track midpoint -- calculated as the midpoint between incoming and outgoing track intersections on the phantom's mid-plane -- serves as a comparative position estimator.

\begin{figure}[!htb]
       \includegraphics[width=0.9\hsize]{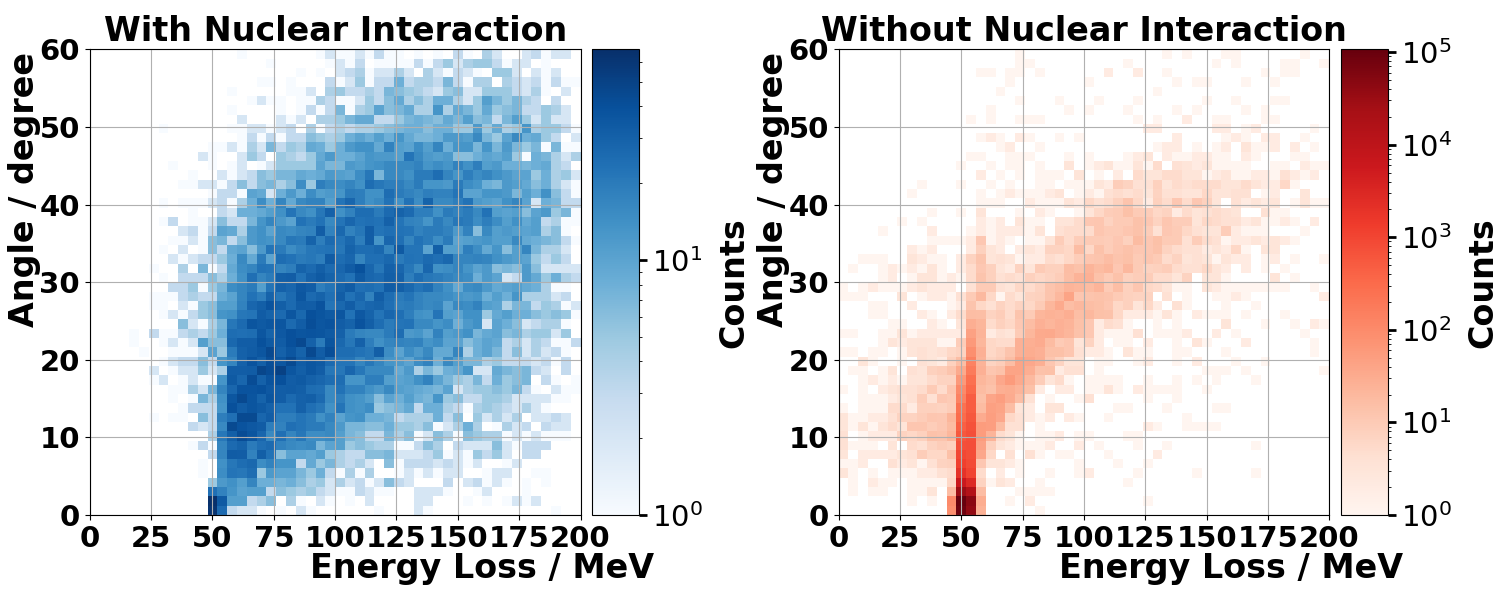}
       
       \includegraphics[width=0.7\hsize]{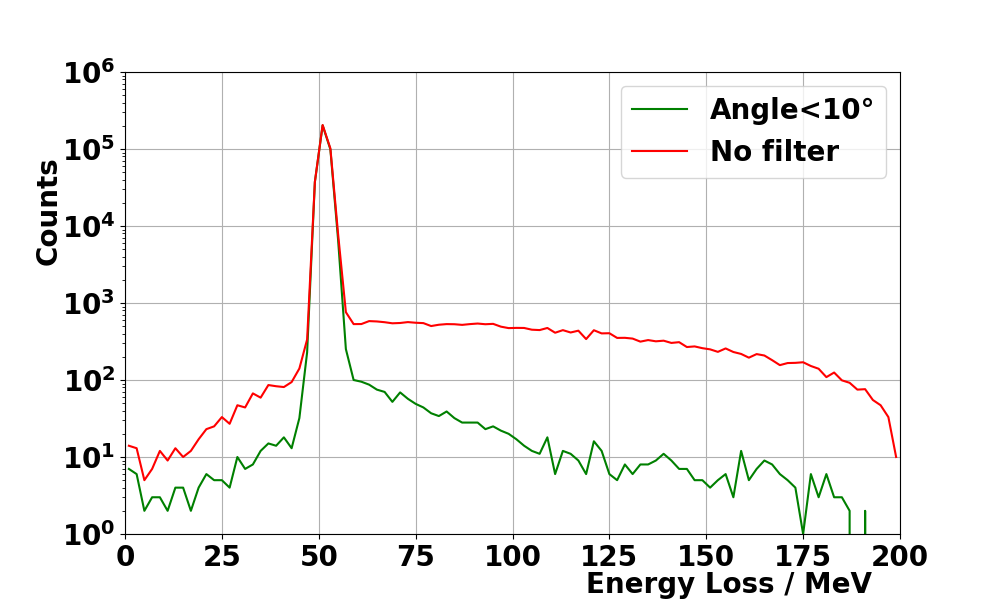}
	\caption{The top figures show energy loss from ground truth data versus $\theta_{\rm s}$ correlation, comparing scenarios with and without nuclear interaction effects.
	The bottom figure shows the distribution of energy loss with and without the filter $\theta_{\rm s}<10^{\circ}$.
	 All figures depict 200~MeV protons traversing a 100~mm water phantom. } 
	\label{fig:Eloss_Tags_RecTracker}
\end{figure}

The proton scattering angle ($\theta_{\rm s}$) constitutes a critical parameter for background suppression in tracking system. 
Fig.~\ref{fig:Eloss_Tags_RecTracker} (top) presents the energy loss versus $\theta_{\rm s}$ correlation for 200~MeV protons traversing a 100~mm water phantom.
While the majority of protons show an expected energy loss near 51~MeV, the subset with higher energy loss exhibits non-negligible angular dispersion. 
This correlation enables effective event filtering based on $\theta_{\rm s}$. 
For example, Fig.~\ref{fig:Eloss_Tags_RecTracker} (bottom) compares the energy loss distributions from ground truth data with and without a $\theta_{\rm s}<10^{\circ}$ filter. 
Applying this filter to only 8\% of proton events reduces the energy loss standard deviation (STD) from 16.7~MeV to 3.3~MeV. 
The selection of specific threshold is detailed in Sec.~\ref{sec:Performance_Track}.

\subsubsection{Residual Range Measurement and Filter on Deposit Energy}
The $\rm n_{PE}$ is measured by SiPMs, producing maps in alternating projections in the range telescope: X-Z planes for even-numbered layers and Y-Z planes for odd-numbered layers. 
Proton track reconstruction is performed using an iterative algorithm:
\begin{enumerate}
    \item \textbf{Seed initialization:} Utilize extrapolated tracks from back tracking stations as initial seeds.
    \item \textbf{Candidate selection and iteration:}
\begin{itemize}
\item Select candidate hits from scintillator bars aligned with or adjacent to current seed
\item Prioritize aligned hit as next-iteration seed. If none exist, choose adjacent hit inclined to the track direction
\item Continuously update the seed and track direction, which are initially derived from extrapolated tracks and refined through iterations
\end{itemize}
    \item \textbf{Termination criteria:} Halt iteration when two consecutive layers within the same projection (X-Z or Y-Z) a lack candidate hit. A single-layer skip tolerance accommodates insensitive regions between scintillator bars.
\end{enumerate}
The reconstructed tracks are primary labeled as "\textbf{perturbed}" tracks if nuclear interaction and large angle scattering (>10$^\circ$) occurs, otherwise as "\textbf{unperturbed}" tracks.
The distribution of deposited energy ($E_{\rm dep}$) is crucial for track discrimination, thereby reducing the uncertainty in reconstructed range.
The conversion from $n_{\rm PE}$ to $E_{\rm dep}$ is calibrated through standalone single-bar simulations, accounting for position-dependent light attenuation along the 128-mm scintillator bar length. 
The hit position along the bar length is determined by analyzing the reconstructed track across adjacent orthogonal layers. 

\begin{figure}[!htb]
	\includegraphics[width=0.45\hsize]{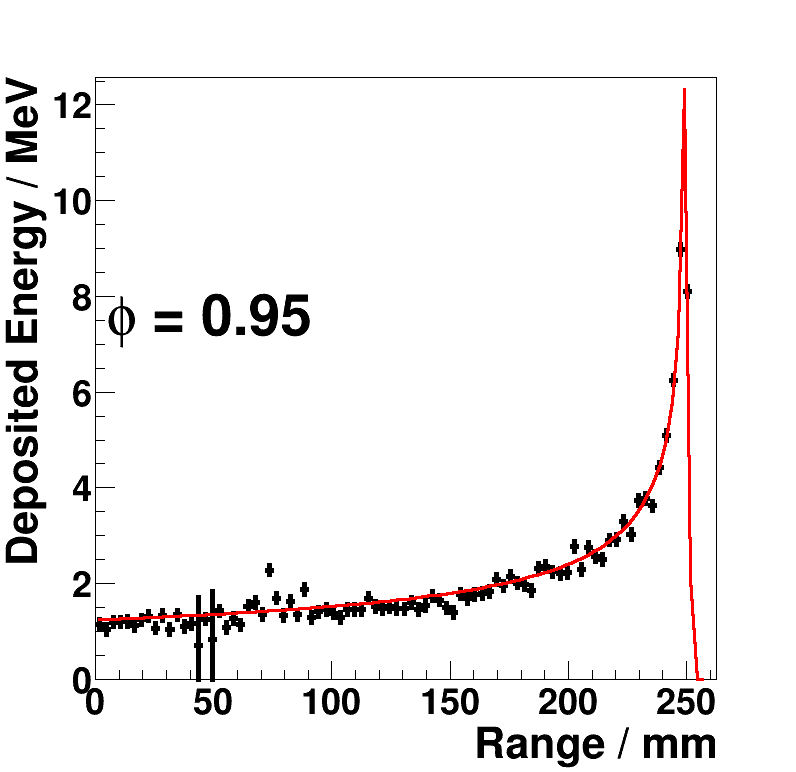}
	\put(-80,40){$(a)$}
	\includegraphics[width=0.45\hsize]{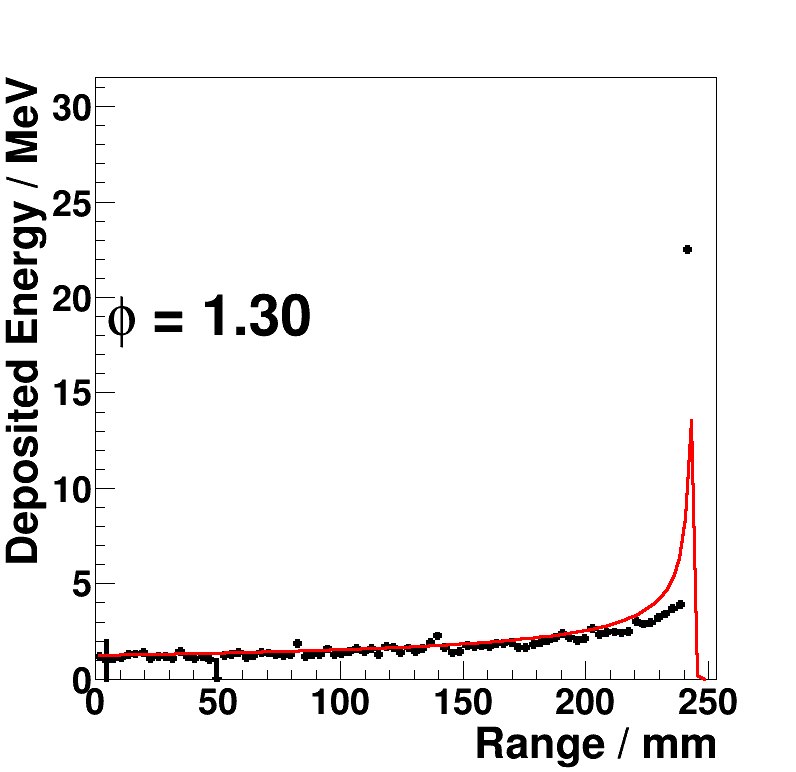}
	\put(-80,40){$(b)$}
	
	\includegraphics[width=0.96\hsize]{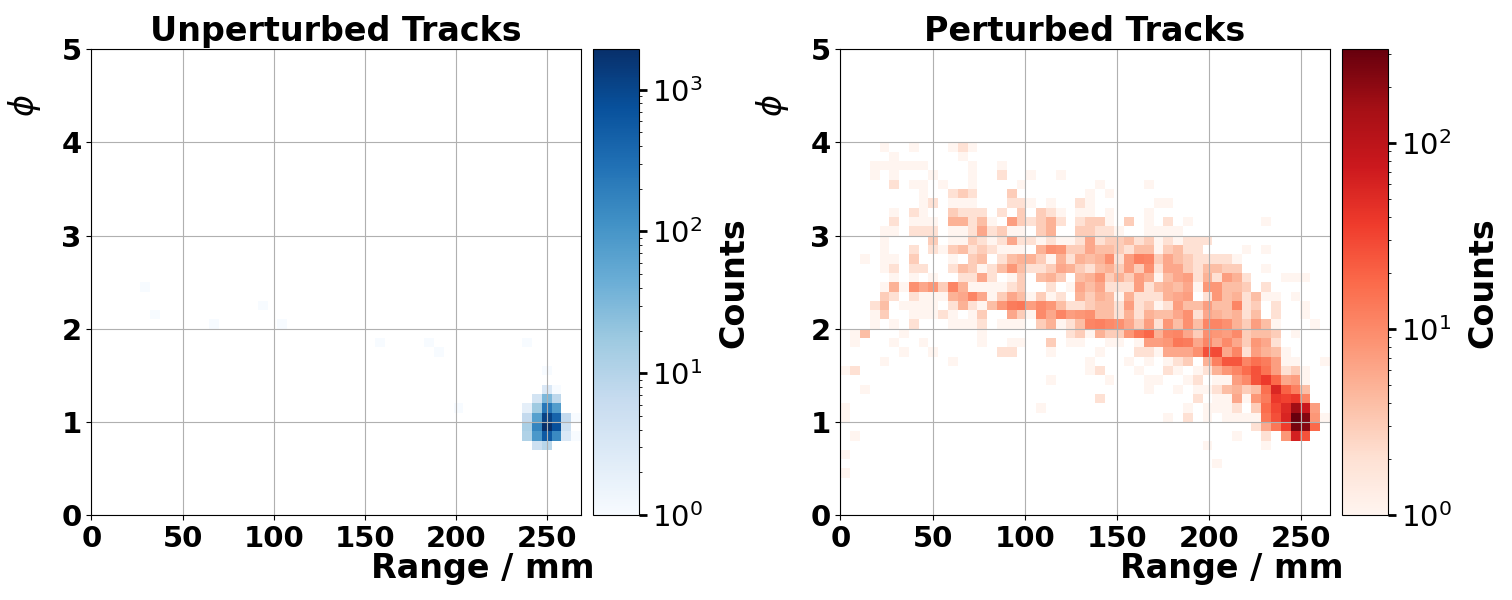}
	\put(-198,30){$(c)$} 
	\put(-88,30){$(d)$} 
	\caption{Bortfeld fit examples and performance for 200~MeV protons without phantom.
	Top figures show Bortfeld fits for (a) \textbf{unperturbed} track and (b) track affected by nuclear interactions, with fixed parameters: \(D_{100} = 14\)~MeV, \(p = 1.77\), \(\sigma = 1\)~mm, \(k = 0.001\).
	Bottom figures are \(\phi\)-range correlation for (c) \textbf{unperturbed} tracks and (d) \textbf{perturbed} tracks.
	} 
	\label{fig:FitCurve_Good_RecRT}  
\end{figure}

The $E_{\rm dep}$ profile is modeled using the Bortfeld function~\cite{ProPhys_1997_BortfeldFunc}, which provides an analytical representation expressed as:
\begin{equation}\label{eq:Bortfeld}
\begin{aligned}
D(z) = &0.65 \cdot D_{100} \cdot \left[ \text{PCF}\left(-\frac{1}{p}, \frac{z-R_0}{\sigma}\right) \right. \\
          & \left. + \sigma \cdot k \cdot \text{PCF}\left(-\frac{1}{p} - 1, \frac{z-R_0}{\sigma}\right) \right]
\end{aligned}
\end{equation}
where
\(D(z)\) is $E_{\rm dep}$ at depth \(z\),
\(D_{100}\) is the peak of $E_{\rm dep}$,
\(p\) is range-energy exponent (material-dependent),
\(R_{0}\) is proton range,
\(\sigma\) is composite Gaussian width accounting for both range straggling and detector resolution,
\(k\) is fluence reduction coefficient,
\(\text{PCF}\) is parabolic cylinder function, implemented from Ref.~\cite{Bortfold_PCF_C}.
The model-dependent parameters (\(D_{100}\), \(p\), \(\sigma\) and \(k\)) are determined by Monte Carlo (MC) study on \textbf{unperturbed} tracks and fixed during track discrimination as shown in Fig.~\ref{fig:FitCurve_Good_RecRT} (a, b), while \(R_{0}\) remains free.
To mitigate effects of energy leakage from insensitive regions, energy deposits below 80\% of the first layer energy (empirically determined) receive increased uncertainty weighting, thereby improving fitting robustness for \textbf{unperturbed} tracks. 

The metric \(\phi = \log_{10}(\chi^{2}/\text{NDF})\) is introduced to quantify goodness of the fit, 
where \(\chi^{2} = \sum_{i=1}^{N_{\text{points}}} \frac{(y_i - f(x_i))^2}{\sigma_i^2}\) represents the discrepancy between the observed values \(y_{i}\) and the fitted values \(f(x_i)\), \(\text{NDF} = N_{\text{layers}} - N_{\text{parameters}}\) is the number of degrees of freedom, \(N_{\text{layers}}\) is the number of data points in the fit and \(N_{\text{parameters}}\) is number of free parameters. 
Fig.~\ref{fig:FitCurve_Good_RecRT} (c, d) shows the \(\phi\)-range correlation for 200 MeV protons without a phantom.
\textbf{Unperturbed} tracks are centered around the target range with \(\phi \approx 1\), while other tracks exhibit increasing \(\phi\) as the range decreases. 
This result provides a basis for selecting tracks based on \(\phi\).

As an alternative track discrimination method, convolutional neural network (CNN) is implemented in the range telescope inspired by the work in Ref.~\cite{CaloRec_2021_BergenpCTCNN}. 
The CNN is constructed using TensorFlow~\cite{Tensorflow_2016} with the following specifications: 
\begin{itemize}
\item Input: 2D projections of the scintillator light yield (\(n_{\text{PE}}\)) in the \(X\text{-}Z\) and \(Y\text{-}Z\) planes. 
\item Output: a scalar probability (0--1) indicating the likelihood of a track being classified as "good".    
\item Architecture: 
The CNN architecture features dual input branches processing two \(n_{\text{PE}}\) images.
Each branch undergoes parallel processing through a convolutional layer (3×3 kernel, ReLU activation), followed by a pooling layer (2×2 pooling) and flatten operation. 
The flattened features from both branches are then concatenated into a unified representation, which feeds into a dense layer (128 units, ReLU activation) before final classification via a softmax output layer. 
\item Training: Adam optimizer~\cite{AdamOptimizer_2014} with a learning rate of \(10^{-3}\), 128 batch size and 10 training epochs, loss function is binary cross-entropy for probabilistic discrimination.  
\item Sample: mixed sample similar to the \textbf{Phantom 2} group, where protons pass a water phantom with thickness ranging from 0 to 150~mm, the thickness step is 1~mm instead of 5~mm, 80\% of the sample is used for training and the rest for testing.
\end{itemize}

Proton tracks are labeled using two complementary methods:  
\begin{itemize}
    \item \textbf{Process tags}: Derived from Geant4  ground truth data. Tracks are labeled as "good" if they are not perturbed by nuclear interaction and large angle scattering (>10$^\circ$).
    \item \textbf{Range tags}: Reconstructed ranges for protons passing a same phantom thickness are fitted with a Gaussian distribution. Tracks within $\mu\pm3\sigma$ are labeled as "good".
\end{itemize}

Process tags are more fundamental but can only be accessed through simulation, so it requires MC-data agreement in \(n_{\text{PE}}\) distribution. 
Range tags can be realized in practical experiment but lack physical interpretability.  

\subsubsection{WEPL Calibration} 
\begin{figure}[!htb]
	\includegraphics[width=0.48\hsize]{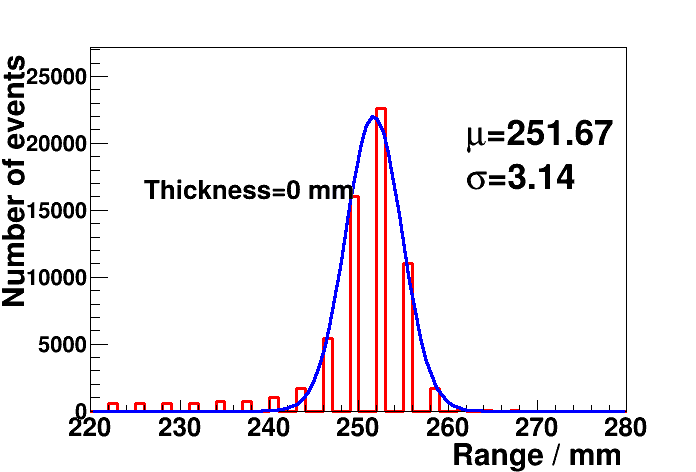}
	\put(-100,23){$(a)$}
	\includegraphics[width=0.48\hsize]{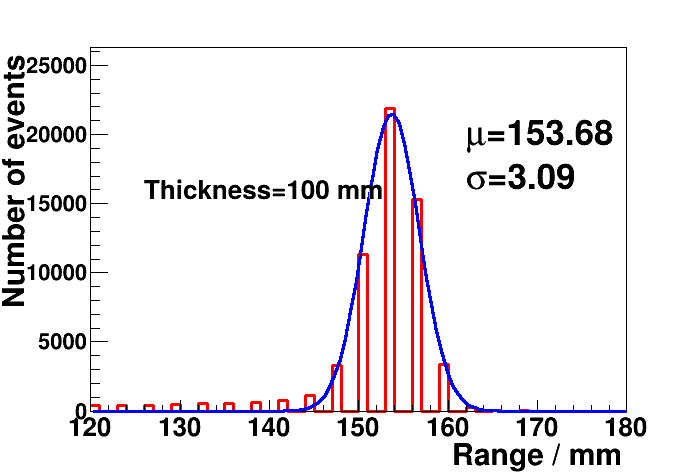}
	\put(-100,23){$(b)$}
	
	\includegraphics[width=0.7\hsize]{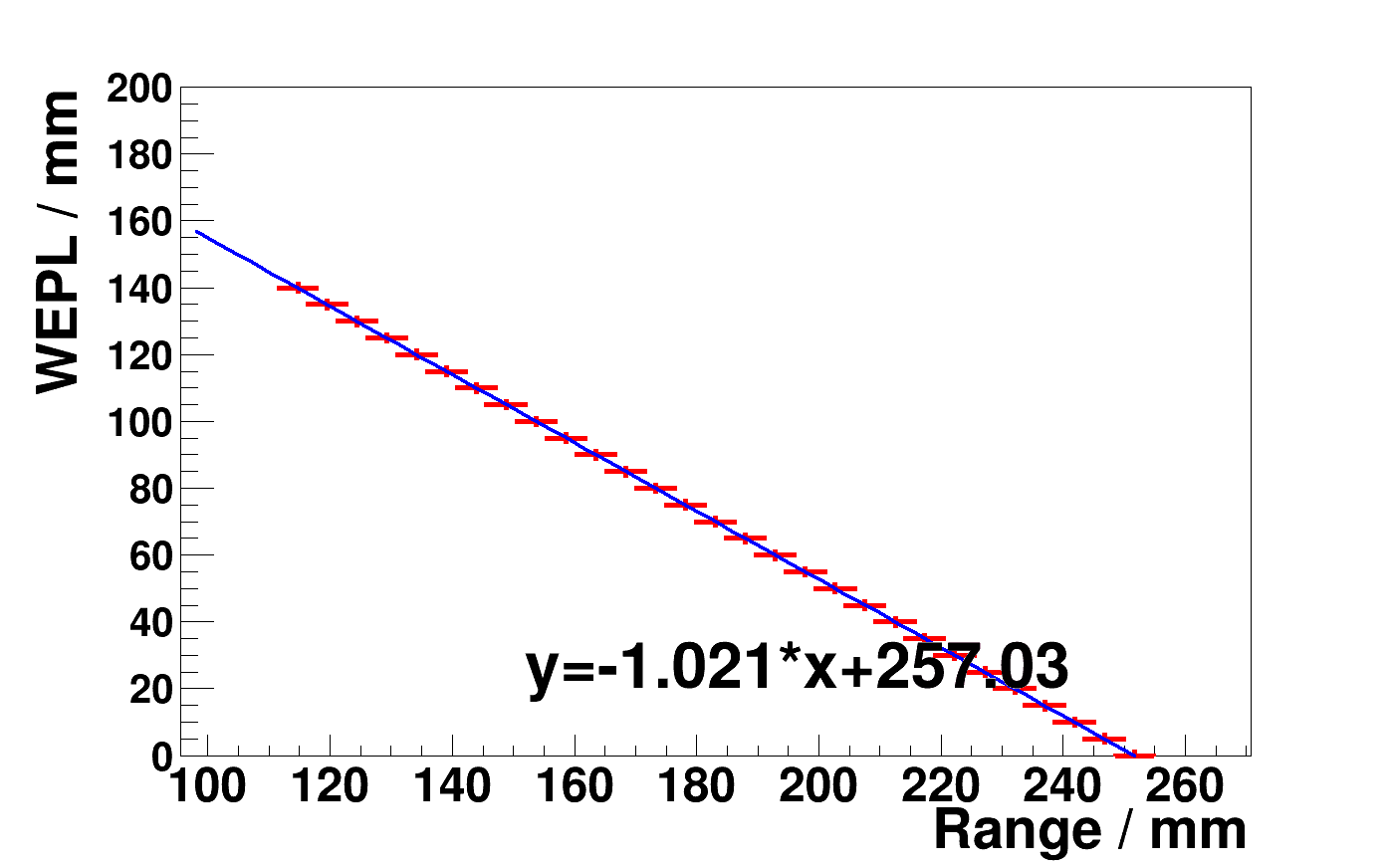}
	\put(-140,23){$(c)$}
	\caption{
	Calibration process and results.
	Top figures are Gaussian fits on the distributions of reconstructed range with phantom thickness of (a) 0 and (b) 100~mm.
	The bottom figure shows the linear fit of the WEPL-range correlation.} 
	\label{fig:RecRT_Calib}  
\end{figure}

The reconstructed residual range is converted to WEPL through a calibration process.
The cuboid phantom group (\textbf{Phantom 2}) with thickness step of 5~mm is utilized to allow protons to pass through varying thicknesses of the phantom.
The reconstructed range distributions for 0~mm (no phantom) and 100~mm thicknesses are fitted with Gaussian functions as shown in Fig.~\ref{fig:RecRT_Calib} (a) and (b).
The means and widths of the Gaussian fits are plotted against the thicknesses, i.e., the WEPL values, and are fitted with a linear function, as illustrated in Fig.~\ref{fig:RecRT_Calib} (c). 

\subsubsection{pCT Imaging}
The protons are grouped into 0.5~mm~$\times$~0.5~mm pixels to construct the WEPL maps.
Each reconstructed proton track provides a WEPL measurement at a spatial position determined by the MLP method, where the WEPL value is determined from the reconstructed proton range and the position is calculated via the MLP method.
Following previous pCT studies~\cite{PhaseII_2016_TechAndFirstExp}, a "pixel-wise" filtering is applied under the assumption of uniform WEPL within each pixel.
The mean and standard deviation ($\sigma$) of WEPL values within each pixel are calculated, and WEPL measurements deviating by more than 2$\sigma$ from the mean are discarded.

The WEPL maps acquired at different angular projections are systematically organized into sinograms. 
Subsequently, the three-dimensional RSP map is reconstructed through FBP utilizing a Hann filter for frequency-domain noise suppression.
In this preliminary investigation, advanced CT reconstruction algorithms are intentionally avoided to specifically isolate and evaluate the fundamental performance characteristics of both the detector system and reconstruction methodology.

%=======================
%   Results
%=======================
\section{Results}
\subsection{Performance of Track Reconstruction}
\label{sec:Performance_Track}
The $\theta_{\rm s}$ and $E_{\rm dep}$ thresholds are determined by balancing the reconstructed WEPL STD and selection rate - two critical metrics for imaging noise reduction.
Optimized thresholds yield selection rates of 97\% for the $\theta_{\rm s}$ filter and 73\% for the $E_{\rm dep}$ filter.

\begin{figure}[!htb]
	\includegraphics[width=0.8\hsize]{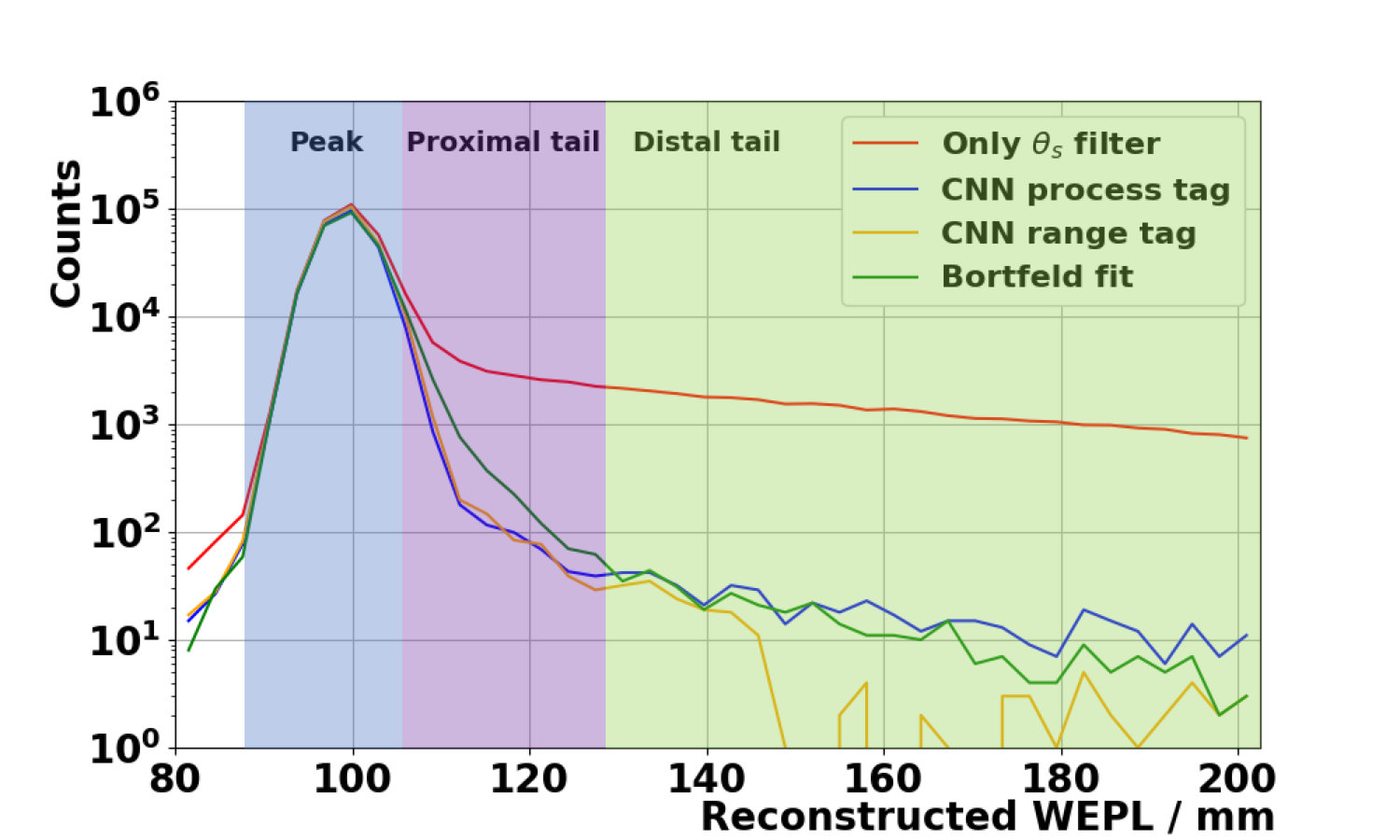}
	\put(-160,20){$(a)$}
	
	\includegraphics[width=0.8\hsize]{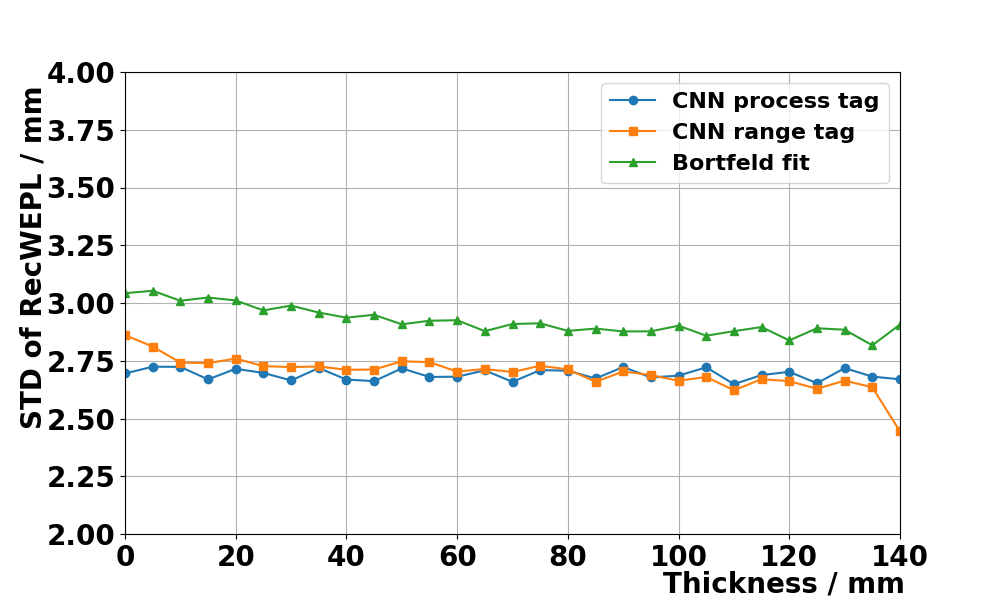}
	\put(-160,20){$(b)$}
	
	\includegraphics[width=0.8\hsize]{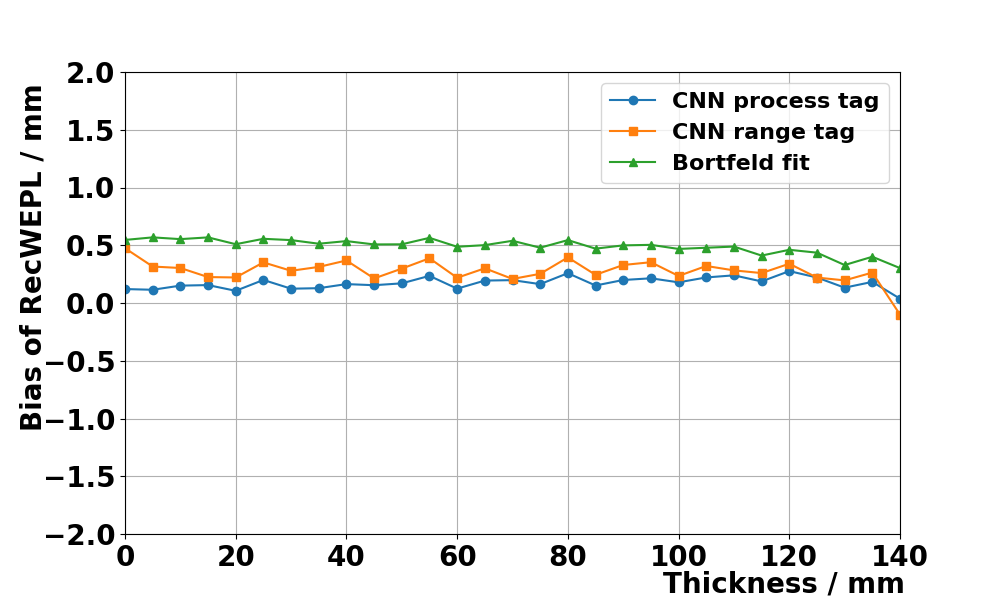}
	\put(-160,20){$(c)$}
	\caption{ 
	Performance comparison of $E_{\rm dep}$ filters: process-tagged CNN, range-tagged CNN, and Bortfeld fitting.
	(a) Reconstructed WEPL distributions after 100 mm-thick phantom. 
	(b) Thickness-dependent WEPL STDs with additional pixel-wise filter ($\mu$=56).
	(c) Thickness-dependent WEPL deviations with additional pixel-wise filter ($\mu$=56).
	}
	\label{fig:RecWEPL_Cuts_Result}
\end{figure}

Fig.~\ref{fig:RecWEPL_Cuts_Result} (a) compares the reconstructed WEPL distributions under $E_{\rm dep}$ filters after applying the $\theta_{\rm s}$ filter.
The WEPL profile is segmented into three regions: the peak, proximal tail and distal tail.
While the $\theta_{\rm s}$ filter marginally reduces background in the tails, all $E_{\rm dep}$ filters achieve significant suppression (1--2 order of magnitude) in both tails. 
Proximal tail events consist of proton tracks undergoing large-angle scattering or nuclear interaction, so CNN filters outperform Bortfeld fitting by leveraging additional scattering discrimination from $n_{\rm PE}$ maps. 
Conversely, distal tail events are governed by nuclear interaction with anomalous $E_{\rm dep}$ distributions, leading to comparable performance between CNN filters and the Bortfeld fitting. 
The range-tagged CNN exhibits superior performance in the WEPL$>150$~mm region, likely due to the training dataset limitations (0--150 mm WEPL range) . 
 
To quantify pixel-wise filter performance, WEPL measurements from same thicknesses are virtually grouped with Poisson-sampled group sizes ($\mu$=56, see Sec.~\ref{sec:Performance_pCT_Imaging}). 
Fig.~\ref{fig:RecWEPL_Cuts_Result} (b) displays the STDs of track-level WEPL after applying the pixel-wise filter. 
The STDs are suppressed to 3~mm-level, showing the effectiveness of track discrimination.
While pixel-wise filters effectively suppress distal tail events, their limited efficacy in proximal tails results in similar STD performance for process-tagged and range-tagged CNNs, whereas Bortfeld fitting exhibits an 8\% higher STD. 
The range-tagged CNN shows anomalies at 0~mm and 140~mm thicknesses, attributable to training set truncation (0--150~mm range).

Although proton energy is not directly reconstructed due to the WEPL-range calibration, its resolution can be derived from the range resolution via:
\begin{equation}
\frac{\sigma_E}{E} = \frac{1}{p} \cdot \frac{\sigma_R}{R}
\end{equation}
where $\sigma_R$=3~mm (estimated from WEPL STD), $R$=260~mm (200 MeV proton range in water), and $p\approx$1.77 (Geiger exponent)~\cite{ProPhys_1997_BortfeldFunc}. 
This yields a relative energy resolution of 0.6\% for 200~MeV protons.

Fig.~\ref{fig:RecWEPL_Cuts_Result} (c) presents the deviation between reconstructed and true WEPL values under combined $\theta_{\rm s}$, $E_{\rm dep}$ and pixel-wise filter. 
Systematic biases are observed: 0.2~mm (process-tagged CNN), 0.3~mm (range-tagged CNN) and 0.5~mm (Bortfeld fitting). 
This bias originates from differences in range determination during calibration versus reconstruction. 
During calibration, the range for a given WEPL is defined as the peak position in the Gaussian-fitted range distribution.
In contrast, reconstructed ranges represent filtered mean values, where the residual tail slightly reduce the mean of reconstructed range, thereby overestimating WEPL.
Crucially, the bias magnitude remains consistent across phantom thicknesses, confirming its independence from phantom geometry. 
This thickness-independent bias confirms the filter's stability. 
Thus, a fixed calibration offset (equal to the negative bias) is applied for each method.

In evaluating $E_{\rm dep}$ filter methods, the process-tagged CNN achieves the lowest STD/bias with negligible phantom thickness dependence, serving as a theoretical benchmark for detector performance limits due to its geometry independence. 
However, its experimental infeasibility restricts practical use unless $n_{\rm PE}$ consistency between data and MC proved.
The range-tagged CNN, while experimentally viable, exhibits significant sample-dependent variations (e.g., WEPL range and step). 
To mitigate this instability, Bortfeld fitting provides a physics-constrained parameterized alternative with enhanced robustness.
For real-world applications, the latter two methods will be experimentally validated to balance accuracy and feasibility.

%=======================
%   pCT imaging
%=======================
\subsection{Performance of Benchmark RSP Imaging}
\label{sec:Performance_pCT_Imaging}
Benchmark pCT imaging employs $4~\times~10^{8}$ protons with 180 projections at 1$^\circ$ angular intervals, achieving 0.5~$\times$~0.5~$\times$~0.5~mm$^{3}$ pixel resolution over a 100~$\times$~100~mm$^{2}$ field. 
The simulated radiation dose is approximately 3.2~mGy.
Assuming uniform proton distribution, this configuration yields an average of $\mu=56$ protons per pixel before filtering and 40 after filtering. 
Pixel-wise WEPL standard deviations below 0.5~mm are achieved for phantom thicknesses of 0--140~mm, exhibiting agreement with the theoretical relationship $\sigma_{\rm pixel}=\sigma_{\rm proton}/\sqrt{\mu}$.

\begin{figure}[!htb]
	\includegraphics[width=0.6\hsize]{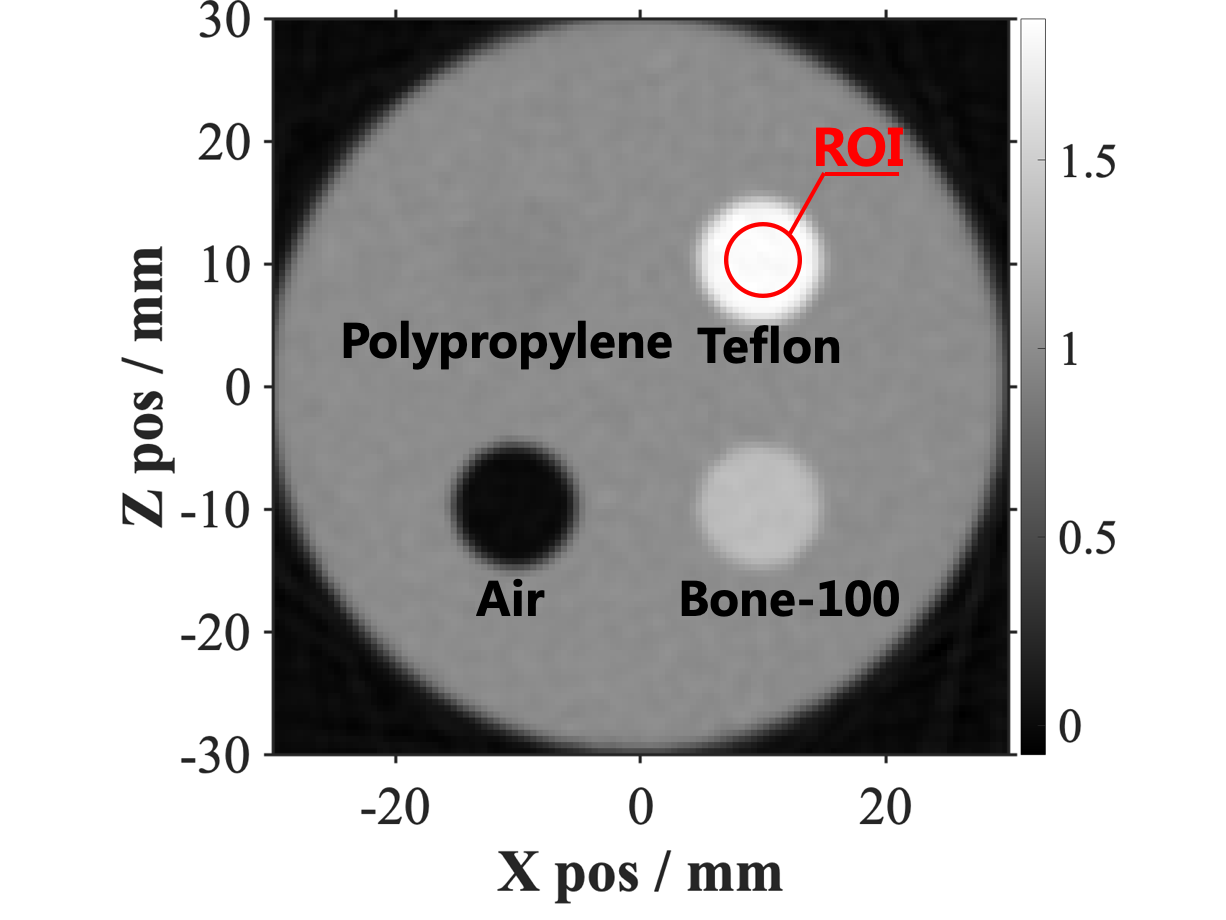}
	\put(-75,-8){$(a)$}

	\includegraphics[width=0.45\hsize]{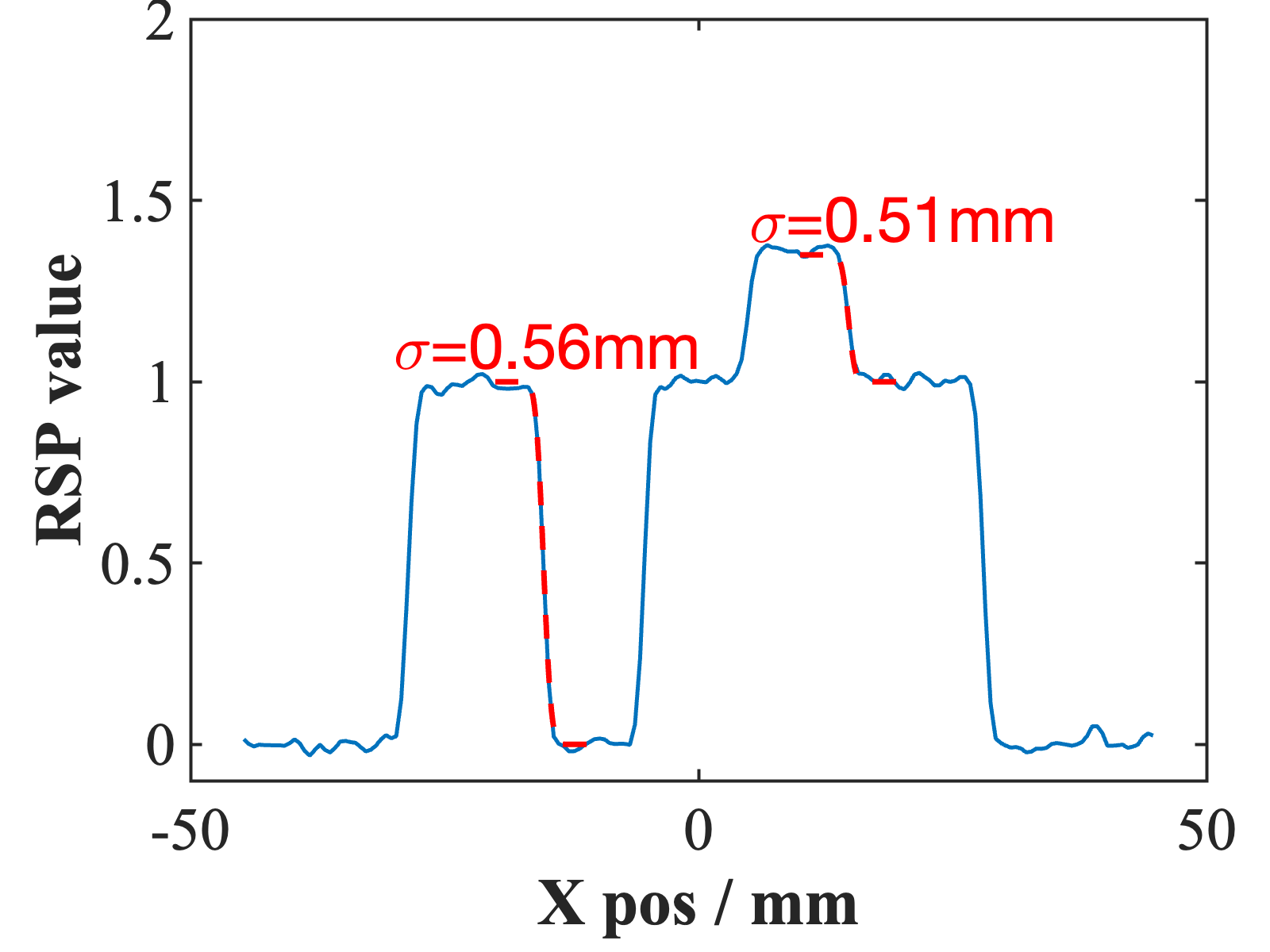}
	\put(-50,-8){$(b)$}
	\includegraphics[width=0.45\hsize]{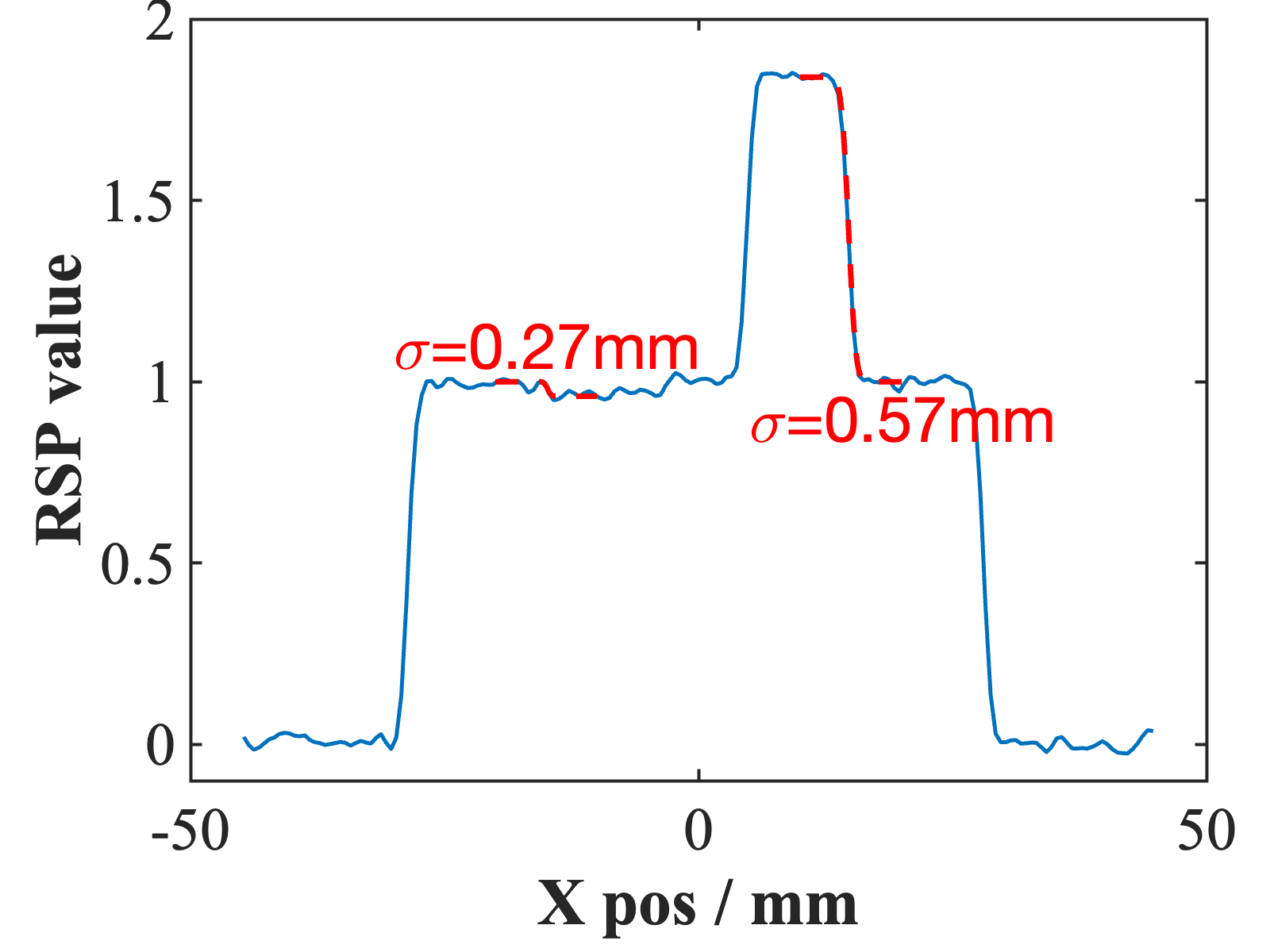}
	\put(-50,-8){$(c)$}
	\caption{
	pCT imaging from $4~\times~10^{8}$ protons with 180 projections and 0.5~$\times$~0.5~$\times$~0.5~mm$^{3}$ pixel resolution.
	Insert coordinates: PP (+10~mm, +10~mm), Teflon (+10~mm, -10~mm), air (-10~mm, +10~mm), and bone (-10~mm, -10~mm). 
	(a) A slice of reconstructed RSP image on the plane where y=0. 
	(b) Profile of RSP distribution along Z=-10~mm with fit on air-water and bone-water boundaries. 
	(c) Profile of RSP distribution along Z=+10~mm with fit on PP-water and Teflon-water boundaries.
	}
	\label{fig:pCT_image}
\end{figure}

The pCT scan is performed on \textbf{Phantom 1}, which includes material inserts comprising PP, Teflon, air and bone-100.
The reconstructed images for a slice are illustrated in Fig.~\ref{fig:pCT_image}.

To quantify the accuracy of the RSP map, cylindrical regions of interest (ROIs) are defined for each insert.
Each ROI is dimensioned to half the radius and height of its corresponding insert.
For volumetric analysis, each ROI spanned 40 consecutive slices, with 80 pixels per slice, yielding a total of 3200 pixels (denoted as $\rm n_{pixel}$).
The mean RSP ($\rm RSP_{mean}$) and width ($\sigma_{\rm pixel,RSP}$) within each ROI are obtained by Gaussian fitting.
The RSP accuracy, representing the systematic deviation from the reference value, is calculated in a same way with Ref.~\cite{PhaseII_2019_CompDECT}:
\begin{equation}\label{eq:RSPAcc}
\begin{aligned}
\rm RSP_{acc} = 100\cdot\frac{RSP_{mean}-RSP_{ref}}{RSP_{ref}}\%
\end{aligned}
\end{equation}
where $\rm RSP_{ref}$ denotes the ground-true RSP value of each material.

Additionally, $\sigma_{\rm pixel,RSP}$ characterizes the noise of spatial heterogeneity values at pixel level.
To estimate the statistical precision of the RSP measurement, the noise is derived by normalizing $\sigma_{\rm RSP}$ with the square root of the total sampled pixels: 
\begin{equation}
\rm RSP_{noise} = \frac{\sigma_{\rm pixel,RSP}}{\sqrt{n_{pixel}}}
\end{equation}

\begin{table*}[htbp]
\caption{
Performance metrics of pCT imaging: reference RSP, accuracy $\pm$ uncertainty, pixel-level variations and spatial resolution under conditions: $4~\times~10^{8}$ protons, 180 projections, 0.5~$\times$~0.5~$\times$~0.5~mm$^{3}$ pixel size and ROI includes 3200 pixels for each insert.
}
\renewcommand{\arraystretch}{1.3}
\begin{tabular}{l|l|c|c|c|c}
\hline 
Method              &Insert             & ~~RSP$_{\rm ref}$~~ & ~~RSP$_{\rm acc}$ (\%)~~      & ~~$\sigma_{\rm pixel,RSP}$~~ & ~~Spatial resolution (mm)~~ \\ \hline 
\multirow{4}{*}{CNN process tag} &Polypropylene &    0.956   & 0.02$\pm$0.03      &           0.013          &       0.46       \\ 
                        &Teflon            &   1.839                     &0.40$\pm$0.01       &            0.014          &      0.54        \\  
                        &Air                 &    0.001                    &-750$\pm$23            &            0.013          &      0.52        \\ 
                        &Bone              &    1.355                    & 0.36$\pm$0.02       &            0.013          &      0.51        \\\hline
\multirow{4}{*}{CNN range tag} &Polypropylene     &    0.956   & 0.57$\pm$0.02      &           0.013          &       0.27       \\ 
                        &Teflon            &   1.839                      &0.15$\pm$0.01        &            0.013          &      0.56        \\  
                        &Air                 &    0.001                     &-796$\pm$24            &            0.013          &   0.57        \\ 
                        &Bone              &    1.355                     & 0.55$\pm$0.02      &            0.013          &      0.51        \\\hline
\multirow{4}{*}{Bortfeld fitting}       &Polypropylene     &    0.956   & 0.45$\pm$0.03      &           0.014          &       0.81       \\ 
                        &Teflon            &   1.839                      &0.36$\pm$0.01        &            0.015          &      0.53        \\  
                        &Air                 &    0.001                     &137$\pm$25            &            0.014          &      0.54        \\ 
                        &Bone              &    1.355                     &0.45$\pm$0.02      &            0.014          &      0.53        \\\hline
\end{tabular}
\label{tab:pCTResults_400M}
\end{table*}

The spatial resolutions are quantified by extracting RSP values near insert-water boundaries, following fits with a step function convolved with a Gaussian function. 
All materials demonstrate spatial resolution of $\sim$0.5~mm except for PP, which exhibits measurement instability due to RSP fluctuations, stemming from its near-water equivalence with only 4\% RSP difference from water.
When excluding the unstable PP results, the MLP method demonstrates approximately 10\% performance enhancement over inter-track midpoint estimation described in Sec.~\ref{sec:Rec_TrackingStation}. 
This limited improvement can be attributed to the fundamental spatial resolution limit imposed by the 0.5~mm pixel size inherently constrains potential improvements.

The pCT imaging results are summarized in Table~\ref{tab:pCTResults_400M}. 
RSP accuracies remain below 0.5\% for all materials except air, where the small absolute RSP value leads to an amplified relative error. 
Pixel-level RSP fluctuations $\sigma_{\rm pixel,RSP}$ show consistency across all inserts, maintaining approximately 0.013. 

\subsection{Performance of Low-Dose RSP Imaging}
Track discrimination demonstrates a significant advantage in statistical efficiency by implementing preliminary suppression of the WEPL tail distribution. 
This preprocessing step significantly reduces the proton density requirements for subsequent pixel-wise filtering.
From study of virtually grouping of WEPL measurements, a sub-3~mm track-level WEPL STD is achieved with a remarkably low proton density of 11 protons per pixel.
So a low-dose proton imaging protocol is proposed with only $2~\times~10^{7}$ protons distributed across 180 projection angles, corresponding to a total absorbed dose of 0.16~mGy. 
The imaging configuration adopts an enlarged pixel size of 1.0~$\times$~1.0~$\times$~1.0~mm$^{3}$, with each ROI containing 400 pixels.

\begin{figure}[!htb]
	\includegraphics[width=0.45\hsize]{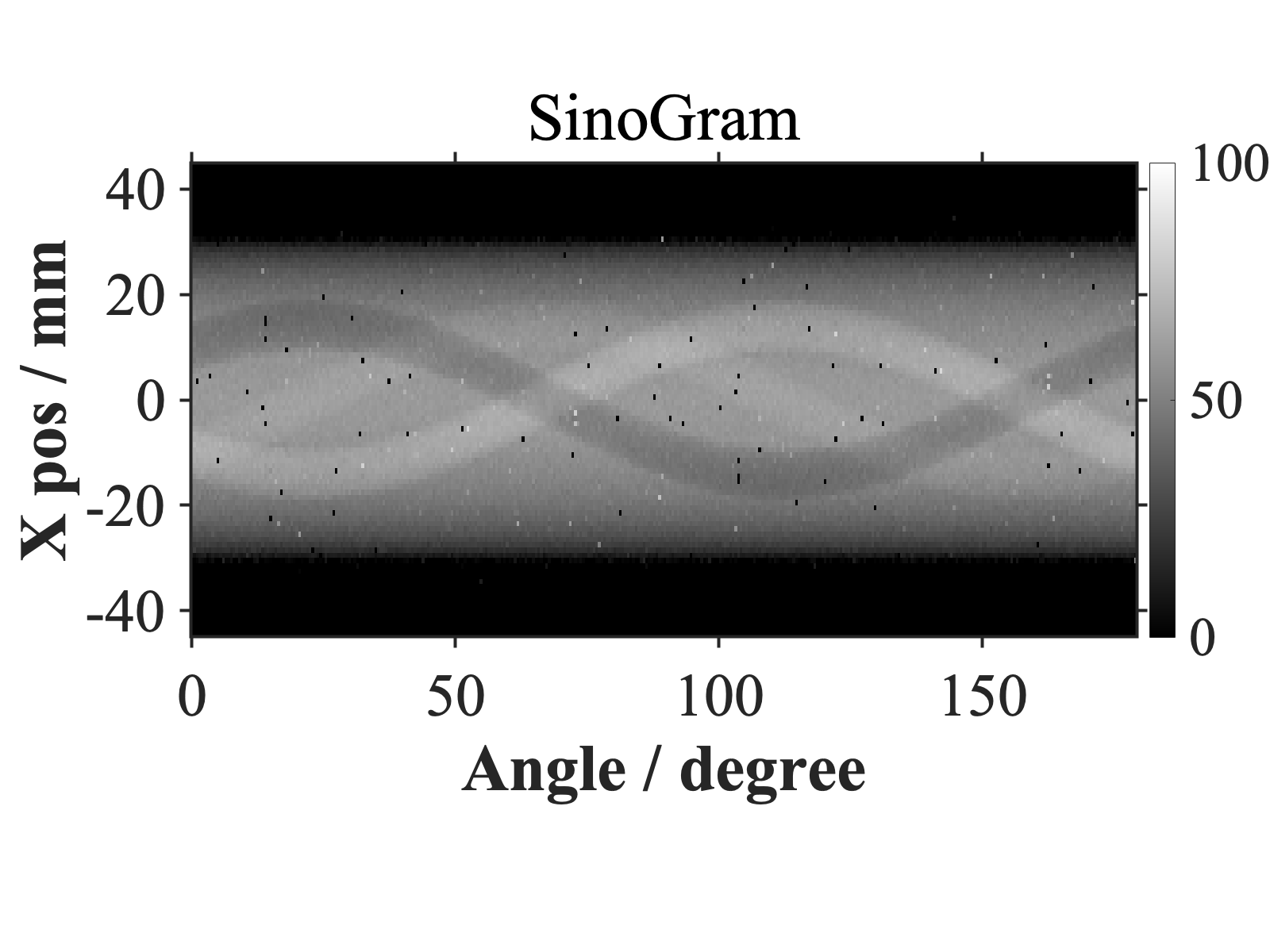}
	\includegraphics[width=0.45\hsize]{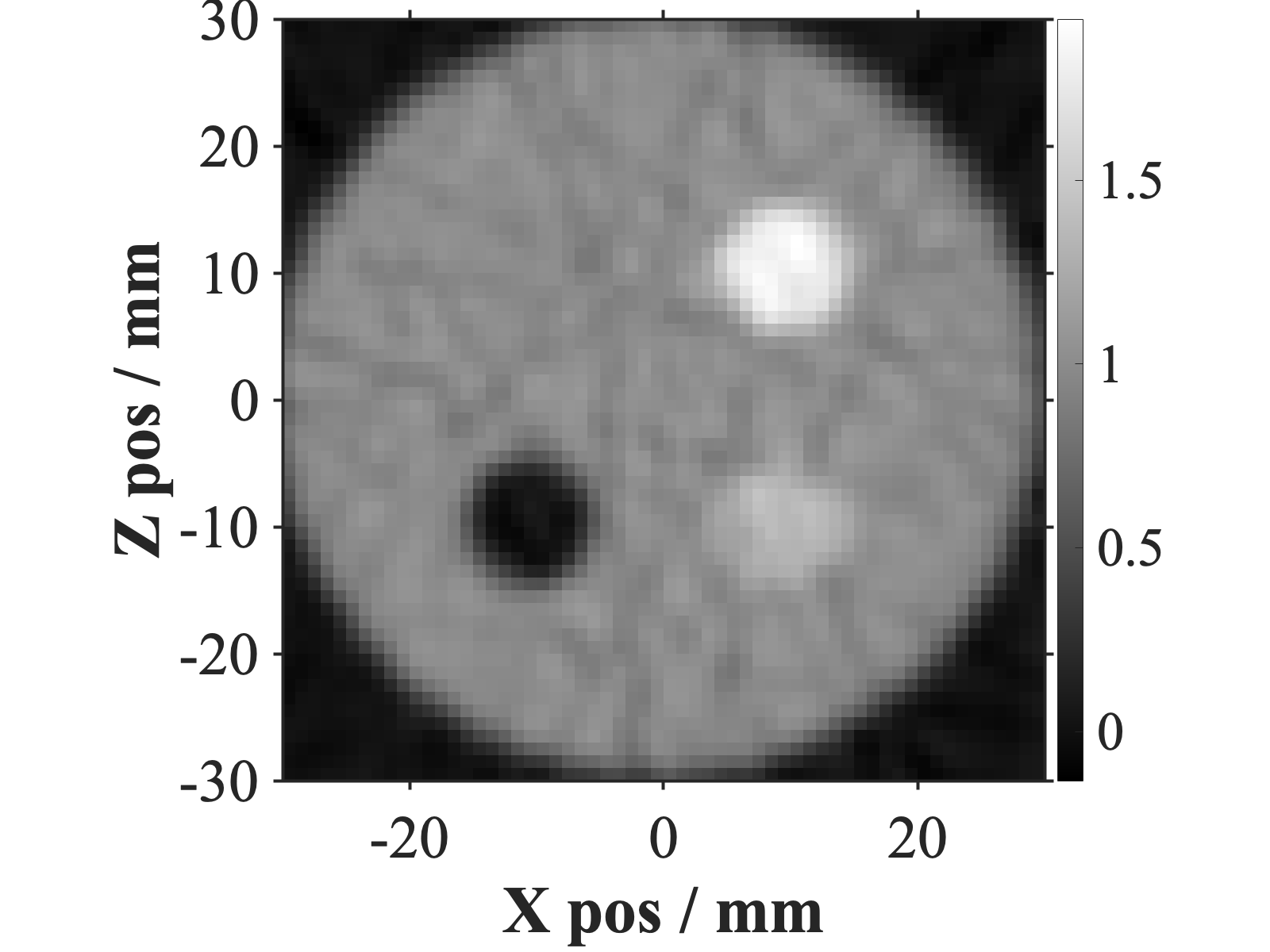}
	\put(-110,10){$(a)$}
	
	\includegraphics[width=0.45\hsize]{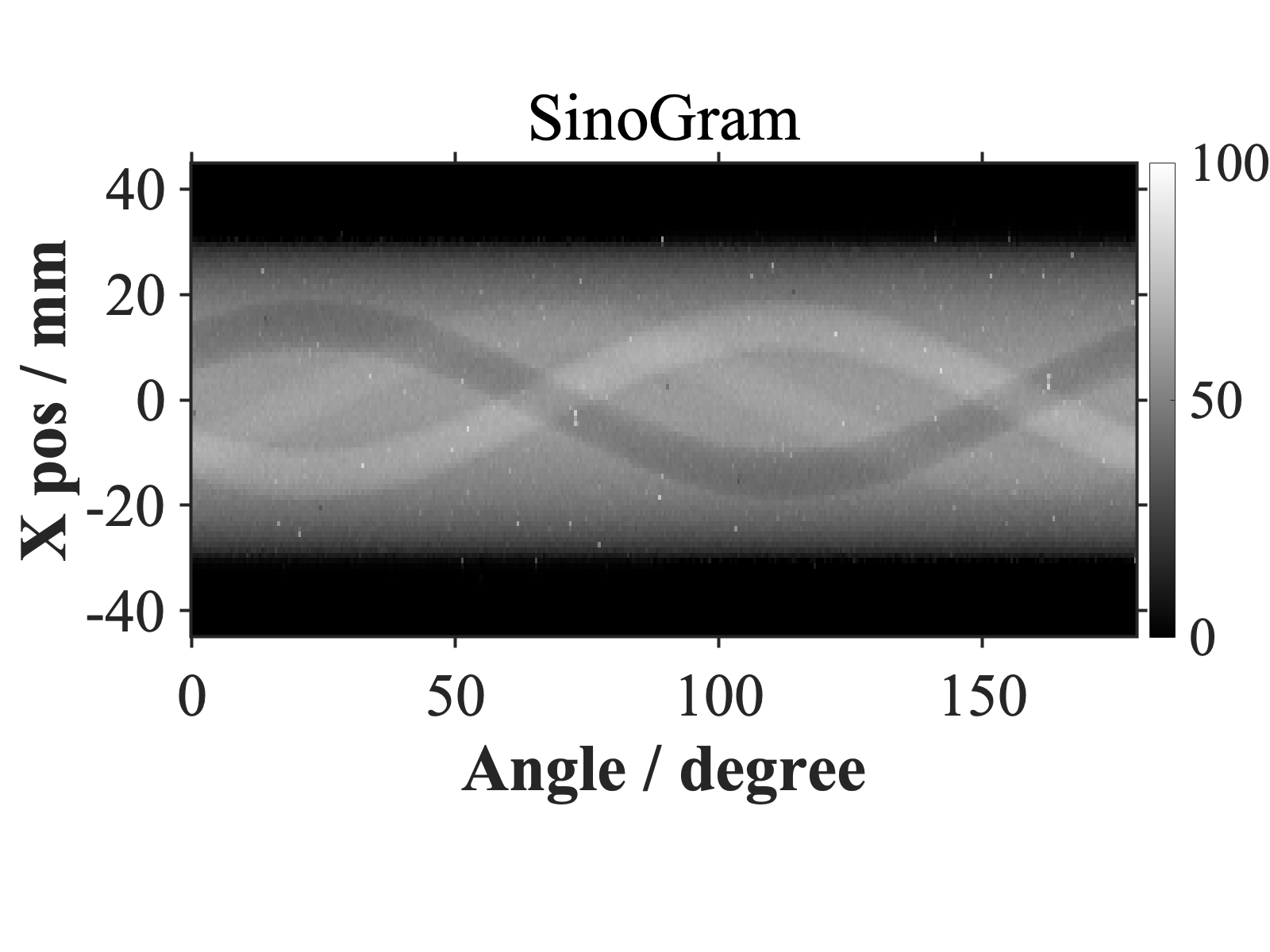}
	\includegraphics[width=0.45\hsize]{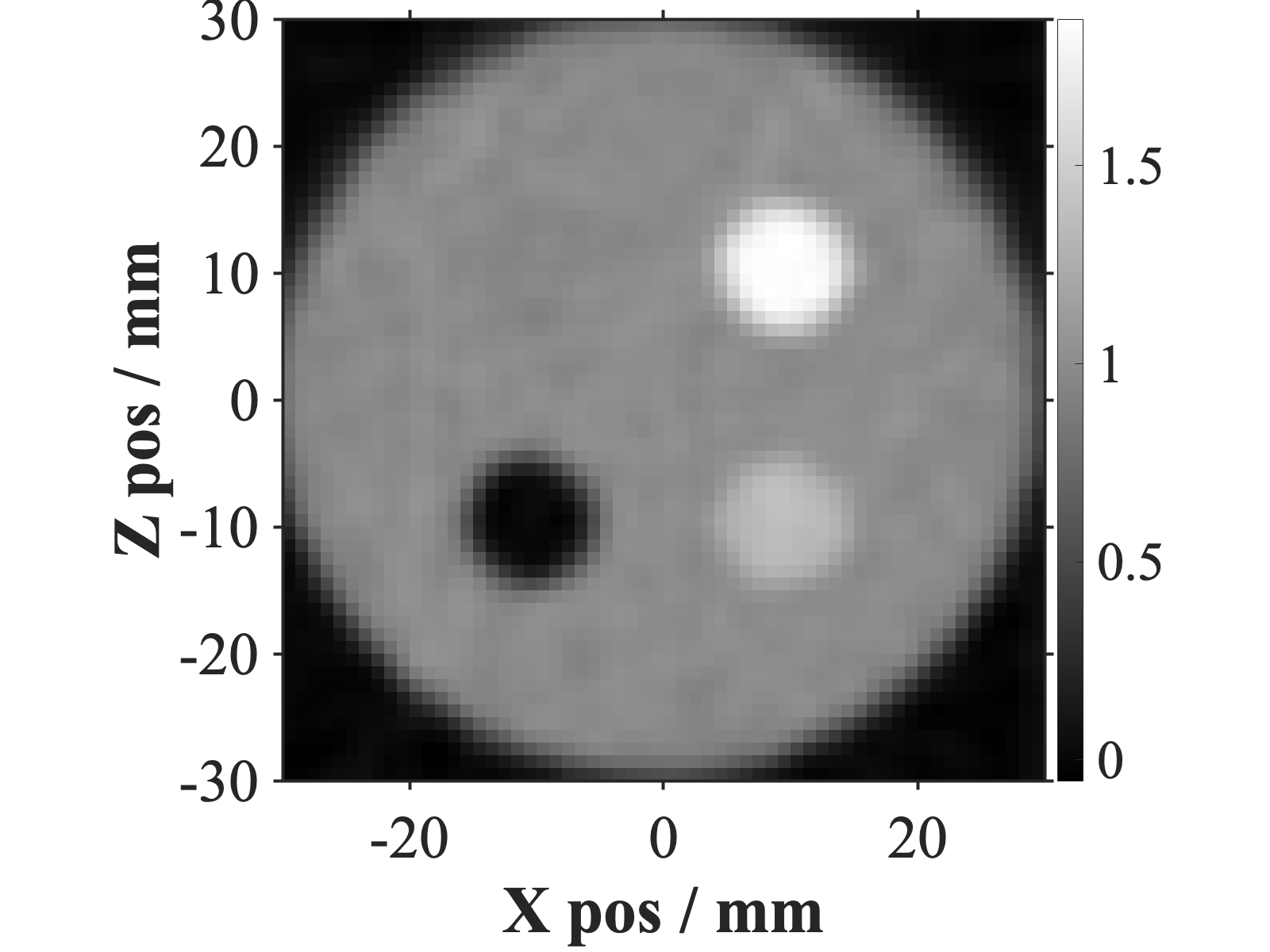}
	\put(-110,10){$(b)$}
	\caption{
	Low-dose pCT imaging from $2~\times~10^{7}$ protons with 180 projections and 1.0~$\times$~1.0~$\times$~1.0~mm$^{3}$ pixel resolution. 
	(a) Original sinogram and RSP image.  
	(b) Artifact-corrected results using proton count thresholding.}
	\label{fig:Image_LowDose}
\end{figure}

The primary reconstructed sinogram and pCT image in Fig.~\ref{fig:Image_LowDose} (a) exhibit linear artifacts caused by noisy pixels in the sinogram where insufficient proton counts compromise the effectiveness of pixel-wise filtering.
To address this issue, a noise correction protocol is implemented in which pixels with $<$5 proton counts are corrected using linear interpolation of adjacent valid pixels ($\ge$5 counts).
This count-driven correction method effectively preserves WEPL accuracy while mitigating artifacts, as demonstrated by the refined sinogram patterns and enhanced pCT reconstruction quality shown in Fig.~\ref{fig:Image_LowDose} (b).
In addition, significant potential exists for developing advanced algorithms leveraging both WEPL data and proton count distributions across neighboring pixels. 

The quantified RSP accuracy and spatial resolution are presented in Table~\ref{tab:pCTResults_20M}. 
While the RSP uncertainty increases by 5-6 times compared to conventional pCT protocols, the RSP accuracy remains <1\% (except for air) while spatial resolution increased to 1.1~mm. 
These results demonstrate the feasibility of clinical pCT implementation at ultralow radiation doses (0.2~mGy level), achieving the dual objectives of <1\% RSP accuracy and millimeter-scale resolution - specifications sufficient for pretreatment positioning in fractionated proton therapy. 

Though exhibiting elevated noise levels, this dose-efficient reconstruction paradigm maintains sufficient imaging fidelity for repeated beam alignment applications where cumulative radiation exposure must be strictly controlled. 
The scanning time is reduced to 2~seconds at a 10~MHz proton detection rate, suggesting the potential of real-time image-guided radiotherapy. 
Imaging quality can be further enhanced through integration with pre-acquired X-ray reference images\cite{CombXray_2023_XrayPriorImage, CombXray_2023_XrayPriorImage2}.

\begin{table*}[htbp]
\caption{
Performance metrics of low-dose pCT imaging with range-tagged CNN: reference RSP, accuracy $\pm$ uncertainty, pixel-level variations and spatial resolution under conditions: $2~\times~10^{7}$ protons, 180 projections, 1.0~$\times$~1.0~$\times$~1.0 mm$^{3}$ pixel size and ROI includes 400 pixels for each insert.}
\renewcommand{\arraystretch}{1.3}
\begin{tabular}{l|c|c|c|c}
\hline 
Insert             & ~~RSP$_{\rm ref}$~~ & ~~RSP$_{\rm acc} (\%)$~~      & ~~$\sigma_{\rm pixel,RSP}$~~ & ~~Spatial resolution (mm)~~ \\ \hline 
Polypropylene &    0.956   & 0.60$\pm$0.13      &           0.024          &       0.99       \\ 
Teflon            &   1.839    &0.01$\pm$0.08        &            0.030          &      1.08        \\  
Air                 &    0.001   &64$\pm$141&            0.028          &       1.09        \\ 
Bone              &    1.355   & 0.40$\pm$0.11      &            0.029          &       1.09        \\\hline
\end{tabular}
\label{tab:pCTResults_20M}
\end{table*}

%=======================
%   Discussion and prospect
%=======================
\section{Discussion and Prospect}
\subsection{Study on Detector Configurations}
The detector configurations are investigated through systematic simulations, with key design parameters compared in Table~\ref{tab:DetConfig}.
The process-tagged CNN is used as $E_{\rm dep}$ filter due to its model-independent tagging. 
The WEPL STDs at the track level serve as the primary evaluation metric. 
Furthermore, reconstructed RSP images demonstrate that $\sigma_{\rm pixel,RSP}$ increases proportionally with WEPL variations.
Notably, all configurations maintain RSP accuracy within 1\% (excluding air) through individualized calibration implemented for each detector setup. 

The range telescope paradigm is designed to relax the requirements for energy resolution of individual scintillator channels. 
This approach enables the potential optimization of lower-bit ADCs and reduces data size, thereby improving readout speed and lowering system cost.
In the simulation, the $n_{\rm PE}$ is discretized into $2^{12}$ values to model the 12-bit ADC resolution of the MPT2321.
For this ADC bit-depth study, the digitization resolution was reduced from 12-bit to 10-bit, 8-bit and 4-bit.
Results in Table~\ref{tab:DetConfig} show that reduce the bit depth maintains robust performance in WEPL measurements, with STD varying by less that 2\%.

The scintillator bar thickness is a critical parameter in range telescope design, as it directly determines the WEPL uncertainty, quantified by:
$\sigma_{\rm WEPL}^{2}=k^{2}\cdot\sigma_{\rm range}^{2}=k^{2}\cdot(\sigma_{\rm proton}^{2}+\sigma_{\rm det}^{2})$,
where $k$ denotes the WEPL-range calibration coefficient from Fig.~\ref{fig:RecRT_Calib} (c).
The intrinsic proton range straggling in plastic scintillator contributes $\sigma_{\rm proton}\approx2.5$~mm, 
while the detector uncertainty $\sigma_{\rm det}=\frac{\rm Thickness}{\sqrt{12}}$ arises from geometrical resolution limitations.
Theoretical calculations demonstrate a baseline uncertainty of 2.75~mm at 3~mm thickness. 
Progressive thickness increases to 4, 5, and 6~mm elevate the uncertainty by 4.1\%, 9.1\%, and 15.0\% respectively.
The simulated STDs in Table~\ref{tab:DetConfig} show agreement with theoretical predictions, thereby confirming the analytical model. 
These results establish 3~mm as the optimal scintillator thickness for precision range measurement applications. 

The lateral segmentation is primarily intended to reconstruct multi-proton tracks. 
However, the STD of single-proton discrimination with CNN increases when the scintillator width is enlarged, becoming comparable to that of the Bortfeld fitting method.
This result indicates an additional advantage from lateral segmentation by enabling the distinction of large-angle scattering events. 

The current proton energy of 200~MeV limits the WEPL measurement range to 0--260~mm. 
To accommodate future needs for extended measurement ranges, higher proton energies will be necessary. 
However, as the proton range increases, intrinsic range straggling also grows, following the relation $\sigma_{\rm proton} \approx kR^{m}$~\cite{RangeStraggling_1993}, where $R$ is the proton range, and $k$ and $m$ are material-dependent constants. 
Consequently, increasing the proton energy from 200~MeV to 250/300~MeV results in a significant rise in WEPL STD, as shown in Table~\ref{tab:DetConfig}. 
These findings highlight a fundamental trade-off between WEPL measurement range and its STD when deriving WEPL from proton range.
Although the WEPL STD increases, the RSP accuracy remains below 1\% due to unbiased calibration and reconstruction.
This ensures reliable performance for future extension from head to body scans, despite the associated increase in pixel noise.

\begin{table}[htbp]
\caption{WEPL STD (track-level) growth rate across detector configurations. The STD$_{\rm BM}$=2.72~mm is under benchmark configuration (12-bit digitization for $n_{\rm PE}$, scintillator thickness=3~mm and width=8~mm, proton energy=200~MeV). The STD$_{\rm X}$ is under corresponding configuration.}
\renewcommand{\arraystretch}{1.3}
\begin{tabular}{l|c|c}
\hline 
Configurations         &   Setting & STD$_{\rm X}$/STD$_{\rm BM}$-1 \\ \hline 
\multirow{3}{*}{Digitization of $n_{\rm PE}$} & 10 bit & 0.3\%                \\
                              & 8 bit  & 0.9\%                \\
                              & 4 bit  & 1.7\%                \\ \hline 
\multirow{3}{*}{Scintillator thickness}    & 4 mm   & 4.3\%                \\
                              & 5 mm   & 9.7\%                \\
                              & 6 mm   & 14.0\%               \\ \hline 
\multirow{3}{*}{Scintillator width}        & 16 mm  & 0.6\%               \\
                              & 32 mm  & 3.8\%                \\
                              & 64 mm  & 5.0\%       \\ \hline
\multirow{2}{*}{Proton energy}        & 250 MeV  & 41.8\%               \\
                              & 300 MeV  & 88.3\%       \\ \hline
\end{tabular}
\label{tab:DetConfig}
\end{table}

\subsection{Proof-of-Concept Prototype}
\label{sec:SmallProto}
\begin{figure}[!htb]
	\includegraphics[width=0.95\hsize]{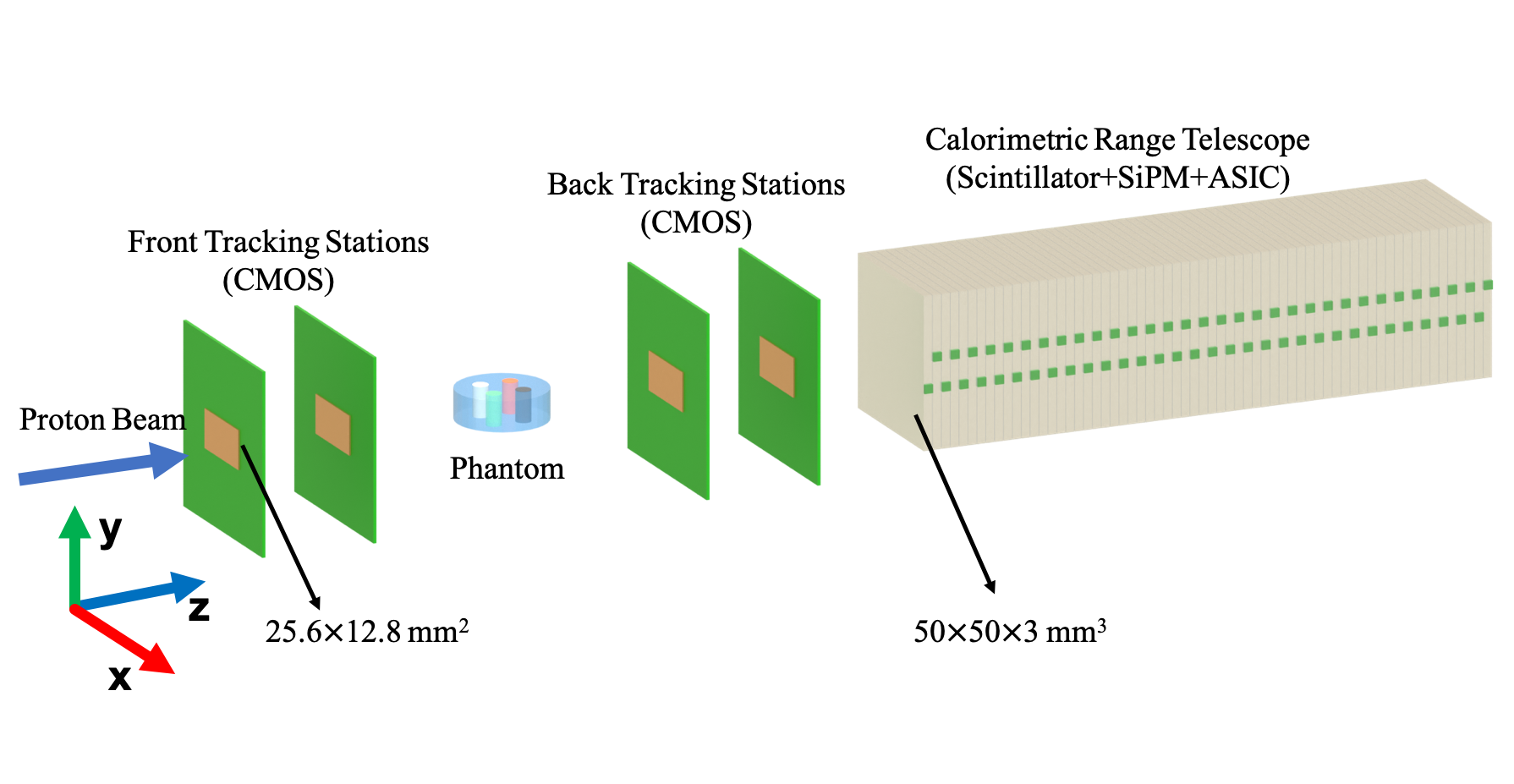} 
	\caption{
	Overview of proof-of-concept prototype structure. Each tracking station consists of a TaichuPix-3 chip (brown) mounted on a readout board (green). 
	The range telescope comprises 64 layers, with each layer containing a scintillator plate (silver) coupled to a SiPM (green).
	}
	\label{fig:pCT_small_structure}
\end{figure}

At the preliminary stage, a simplified prototype is designed to validate the core principles.
The tracking station uses only one TaichuPix-3 chip per detection layer. 
The range telescope comprises 64 layers of 50~$\times$~50~$\times$~3 mm$^{3}$ BC-408 scintillator plates, each coupled to a single SiPM (S13360-3025PE), as shown in Fig.~\ref{fig:pCT_small_structure}. 
To prevent overlap, the SiPMs in adjacent layers are intentionally offset.
This configuration provides a detection area of 25.6~$\times$~12.8~mm$^{2}$ for 100~MeV protons, allowing the reconstruction of only one proton per event. 

A 100~MeV proton beam evaluation is planned for 2025, with this prototype serving for initial system verification and experimental data acquisition. 
Subsequent enhancements will focus on:
(1) Implementation of nonideal simulation and mitigation strategies in reconstruction addressing noise, dead area and crosstalk in both tracking system and range telescope.
(2) Discrimination verification of Bortfeld fitting and range-based CNN with realistic $n_{\rm PE}$ distributions. 
(3) Testing of the real system performance and establishment of the imaging protocols, especially the statistical requirement per pixel.

\subsection{Further Improvement}
Further improvements in proton detection rates primarily depend on multi-proton reconstruction techniques. 
The multi-proton reconstruction algorithm under development for the range telescope aims to discriminate overlapping proton tracks while maintaining the existing single-track discrimination performance. 
By addressing challenges in accurately iterating hits and properly allocating light yield at proton track intersections, this algorithm targets the simultaneous reconstruction of 3--5 protons while maintaining a WEPL STD of $\sim$3~mm. 
Additionally, the tolerance for energy measurement precision (validated via $n_{\rm PE}$ digitization studies) enables potential detection rate improvements by adopting faster SiPMs and readout chips.

This study adopts the simple FBP algorithm for CT reconstruction, with a focus on the track-level implementation. 
For future pCT imaging, we plan to integrate advanced reconstruction techniques such as DROP-TVS~\cite{CTAlg_2010_TVSinpCT}, which offer superior noise reduction .

%=======================
%   Conclusion
%=======================
\section{Conclusion}
This study presents an innovative pCT system that addresses traditional limitations in energy measurement precision and energy-range correlation through a multi-stage filtering strategy integrated into the imaging workflow. 
By sequentially applying scattering angle filtering in the tracking system, energy deposition filtering in the range telescope, and pixel-wise filtering during WEPL image reconstruction, the system achieves remarkable performance: 0.5~mm spatial resolution and <1\% RSP accuracy in simulations using $4\times10^{8}$ protons.
The key innovation lies in this multi-filter architecture's statistics-easing effect, which substantially reduces proton count requirements compared to conventional pixel-wise filtering alone. 
This enables ultra-low-dose imaging feasibility, demonstrating sub-1\% RSP accuracy with just $2\times10^{7}$ protons (0.16~mGy dose) while maintaining 1.1~mm resolution. Such capabilities directly address two critical clinical needs: repeatable patient positioning for fractionated therapy and potential real-time imaging applications.

Systematic design analyses reveal fundamental principles for system optimization: scintillator geometry studies establish thickness and width tolerances for range telescope performance, and digitization investigations demonstrate robustness against energy resolution compromises.

Future work will focus on prototype validation and enhancement strategies, particularly crosstalk and noise mitigation. 
This architecture, combining physical filtering with advanced algorithms, provides a scalable framework for pCT systems that simultaneously achieve high RSP resolution, ultra-low radiation dose, and clinical practicality.
\section{Bibliography}
%\bibliography-style{abbrv}

%\clearpage
\bibliographystyle{unsrtnat}
\bibliography{reference}

\end{document}